\documentclass[npg]{copernicus}

\newcommand{\vc}[1]{\textbf{\em #1}}

\newcommand{\pder}[2]{\frac{\partial #1}{\partial #2}}

\newcommand{\aap}{Astron. Astrophys.}
\newcommand{\an}{Astron. Nach.}
\newcommand{\apj}{Astrophys. J.}
\newcommand{\apjl}{Astrophys. J. Lett.}
\newcommand{\apss}{Astrophys. Space Sci.}
\newcommand{\araa}{Annu. Rev. Astron. Astrophys.}

\newcommand{\eps}{Earth, Planets, and Space}
\newcommand{\fcp}{Fund. Cosm. Phys.}
\newcommand{\grl}{Geophys. Res. Lett.}
\newcommand{\jcp}{J. Comp. Phys.}
\newcommand{\jgr}{J. Geophys. Res.}
\newcommand{\jpcs}{J. Phys. Conf. Ser.}
\newcommand{\jpp}{J. Plas. Phys.}
\newcommand{\jsc}{J. Sci. Comp.}
\newcommand{\lnp}{Lec. Not. Phys.}
\newcommand{\lrsp}{Liv. Rev. Sol. Phys.}
\newcommand{\mnras}{Mon. Not. R. Astron. Soc.}
\newcommand{\nat}{Nature}
\newcommand{\planss}{Plan. Space Sci.}
\newcommand{\pof}{Phys. Fl.}
\newcommand{\pfb}{Phys. Fl. B}
\newcommand{\pop}{Phys. Plas.}
\newcommand{\pra}{Phys. Rev. A}
\newcommand{\prl}{Phys. Rev. Lett.}
\newcommand{\rmp}{Rev. Mod. Phys.}
\newcommand{\rmxac}{Rev. Mex. Astron. Astrofis. Conf. Ser.}
\newcommand{\solphys}{Sol. Phys.}
\newcommand{\ssr}{Space Sc. Rev.}

\begin{document}

\title{Reconnection Studies Under Different Types of Turbulence Driving}
\author[1]{G.~Kowal}
\author[2]{A.~Lazarian}
\author[3]{E.~T.~Vishniac}
\author[4]{K.~Otmianowska-Mazur}
\affil[1]{Instituto de Astronomia, Geof\'\i sica e Ci\^encias Atmosf\'ericas, Universidade de S\~ao Paulo, Rua do Mat\~ao, 1226 -- Cidade Universit\'{a}ria, CEP 05508-090, S\~ao Paulo/SP, Brazil}
\affil[2]{Department of Astronomy, University of Wisconsin, 475 North Charter Street, Madison, WI 53706, USA}
\affil[3]{Department of Physics and Astronomy, McMaster University, 1280 Main Street West, Hamilton, ON L8S 4M1, Canada}
\affil[4]{Obserwatorium Astronomiczne, Uniwersytet Jagiello\'nski, ul. Orla 171, 30-244 Krak\'ow, Poland}

\runningtitle{Reconnection Under Different Turbulence}
\runningauthor{Kowal et al.}

\correspondence{G.~Kowal (kowal@astro.iag.usp.br)}

\maketitle

\begin{abstract}
We study a model of fast magnetic reconnection in the presence of weak
turbulence proposed by \cite{lazarian99} using three-dimensional direct
numerical simulations.  The model has been already successfully tested in
\cite{kowal09} confirming the dependencies of the reconnection speed $V_{rec}$
on the turbulence injection power $P_{inj}$ and the injection scale $l_{inj}$
expressed by a constraint $V_{rec} \sim P_{inj}^{1/2} l_{inj}^{3/4}$ and no
observed dependency on Ohmic resistivity.  In \cite{kowal09}, in order to drive
turbulence, we injected velocity fluctuations in Fourier space with frequencies
concentrated around $k_{inj}=1/l_{inj}$, as described in Alvelius (1999).  In
this paper we extend our previous studies by comparing fast magnetic
reconnection under different mechanisms of turbulence injection by introducing a
new way of turbulence driving.  The new method injects velocity or magnetic
eddies with a specified amplitude and scale in random locations directly in real
space.  We provide exact relations between the eddy parameters and turbulent
power and injection scale.  We performed simulations with new forcing in order
to study turbulent power and injection scale dependencies.  The results show no
discrepancy between models with two different methods of turbulence driving
exposing the same scalings in both cases.  This is in agreement with the
Lazarian \& Vishniac (1999) predictions.  In addition, we performed a series of
models with varying viscosity $\nu$.  Although Lazarian \& Vishniac (1999) do
not provide any prediction for this dependence, we report a weak relation
between the reconnection speed with viscosity, $V_{rec}\sim\nu^{-1/4}$.
\end{abstract}

\introduction

Magnetic fields are observed in many astrophysical objects and usually play an
important or even crucial role in their dynamics \citep[see][e.g.]{crutcher99,
beck02, vallee97, vallee98}.  They are a key ingredient of astrophysical
processes such as magneto-rotational instability, magnetic dynamo, transport and
acceleration of cosmic rays, accretion disks, turbulence, solar phenomena, gamma
ray bursts, etc. \citep{balbus98, parker92, hanasz09, kulpa11, schlickeiser85,
melrose09, elmegreen04, kotera11}.

Magnetic fields are solenoidal and evolve only through changes in the curl of
the electric field.  In the limit of zero resistivity the topology of the field
lines is a constant of motion and the magnetic flux threading any fluid element
is constant.  Generating large scale magnetic fields requires some kind of
battery effect, like the Biermann battery \citep{khanna98} and generating strong
large scale magnetic fields requires a dynamo \cite[see][for example]{parker92}.
 In the limit of very small resistivity, which is typical for astrophysical
objects, the magnetic flux is ``frozen in'' and magnetic field lines will resist
passing through one another or changing their topology \citep{moffat78}.  Due to
the presence of plasma motions, in particular turbulence, this would result in a
very complex tangle of field lines in real objects, with negligible large scale
magnetic flux.  However, observations indicate that the mean and turbulent
components of magnetic fields in many astrophysical objects are of similar
strengths \cite[see][for example]{beck02}.  This implies the existence of a
process which can violate the frozen-in condition on dynamical time scales, i.e.
fast magnetic field reconnection.

The first analytic model for magnetic reconnection was proposed independently by
\citet{parker57} and \citet{sweet58}.  Sweet-Parker reconnection has the virtue
that it relies on a robust and straightforward geometry.  Two regions with
uniform magnetic fields are separated by thin current sheet.  The speed of
reconnection is given roughly by the resistivity divided by the sheet thickness.
 However, the plasma in the current sheet is constrained to move along the local
field lines, and is ejected from the edge of the current sheet at the Alfv\'en
speed, $V_A$.  Since the width of the current sheet limits the flux of expelled
plasma, the overall reconnection speed is reduced from the Alfv\'en speed by the
square root of the Lundquist number, $S\equiv LV_A/\eta$, where $\eta$ is the
resistivity and $L$ is the length of the current sheet.  In most astrophysical
contexts $S$ is very large and the Sweet-Parker reconnection speed,
$V_{SP}\approx V_A S^{-1/2}$, is negligible.  Fast reconnection requires that
the dependence on $\eta$ be erased.  Given the simplicity of the Sweet-Parker
setup, this requires that the simple geometry of the current sheet must be
broken.

The realization that Sweet-Parker reconnection is inadequate to explain magnetic
reconnection in an astrophysical context was immediately apparent, and gave rise
to decades of research on models of fast reconnection \citep[see][for
reviews]{biskamp00,priest00}.  The first proposal was to replace the current
sheet with an X-point configuration, so that the "sheet" thickness and length
are comparable.  This is the basis for Petschek's model of fast reconnection
\citep{petschek64}.  However, a dynamically self-consistent X-point requires
that the outflow prevent a general collapse into a narrow current sheet.
Otherwise we would expect that the same bulk forces that brought the magnetic
field lines together would lead to Sweet-Parker reconnection. \citet{petschek64}
proposed that slow-mode shocks on either side of the X-point would serve this
purpose.  Moreover, those shocks are responsible for converting most of the
magnetic energy into the heat and kinetic energy.  The X-point in this model has
an overall size which depends on resistivity, but since the magnetic field
decrease logarithmically when approaching the current sheet (due to the
assumption of the current-free magnetic field in the inflow region), the
resulting reconnection speed is some fraction of $V_A$.  Numerical simulations
with uniform resistivity \citep{biskamp96} have showed that in the MHD limit the
shocks fade away and the contact region expands into a sheet.  The only way to
make the Petschek configuration stable is by introducing the local nonuniform
resistivity \citep{ugai77,scholer89,ugai92,yan92,forbes01,shibata11}.

This leaves the possibility that X-point reconnection is stable when the plasma
is collisionless.  Numerical simulations \citep{shay98,shay04} have been
encouraging.  However, there are several important issues that remain
unresolved.  First, it is not clear that this kind of fast reconnection persists
on scales greater than the ion inertial scale \cite[see][]{bhattacharjee03}.
Several numerical studies \citep{wang01,smith04,fitzpatrick04} have found large
scale reconnection speeds which depend on resistivity, i.e. are not fast.
Second, in many circumstances the magnetic field geometry does not allow the
formation of X-point reconnection.  For example, a saddle-shaped current sheet
cannot be spontaneously replaced by an X-point.  The energy required to do so is
comparable to the magnetic energy liberated by reconnection, and must be
available beforehand.  Finally, the requirement that reconnection occur in a
collision less plasma restricts this model to a small fraction of astrophysical
applications.  For example, while reconnection in stellar coronae might be
described in this way, stellar chromospheres are not.  More generally
\cite{yamada07} estimated that the scale of the reconnection sheet should not
exceed about 40 times the electron mean free path.  This condition is not
satisfied in many environments which one might naively consider to be
collisionless, among them the interstellar medium.  The conclusion that stellar
interiors and atmospheres, accretion disks, and the interstellar medium in
general does not allow fast reconnection is drastic and unpalatable.

An alternative to the X-point geometry is to consider magnetic fields that are
chaotic, even if only weakly so.  Requiring the plasma to flow along the local
magnetic field implies a powerful constraint on reconnection only if the field
lines themselves are laminar.  \citet[][hereinafter LV99]{lazarian99} proposed a
model for fast reconnection which depends on the presence of turbulence, and its
production of weakly stochastic field lines (also briefly described in
Section~\ref{sec:lv99model}).  Turbulence is a natural consequence of convection
in stars and of the magnetorotational instability in accretion disks \citep[for
a review see][]{balbus98}.  In addition, it is now generally accepted that the
``Big Power Law in the Sky'' indicates the presence of turbulence on scales from
tens of parsecs to thousands of kilometers \citep{armstrong95,chepurnov10a}.
Among other sources, evidence for this comes from studies of atomic hydrogen
spectra in molecular clouds and galaxies \cite[see][see also review by
\citeauthor{lazarian09}, \citeyear{lazarian09} and references
therein]{lazarian99b,stanimirovic01,padoan06,padoan09,chepurnov10b}, as well as
recent studies of emission lines and Faraday rotation
\cite[see][]{burkhart10,gaensler11}.  LV99's model uses the properties of
turbulence to predict broad outflows from extended current sheets.  The
diffusivity of magnetic field line trajectories in a turbulent plasma implies
that flows can follow local magnetic field lines without being confined to the
current sheet.  When the turbulent diffusivity is less than the ohmic
resistivity, this model reduces to the Sweet-Parker reconnection model.

The first test of the LV99 model using three-dimensional (3D) MHD simulations
was performed in \cite{kowal09}.  The main predictions of the model were
confirmed.  In this paper we provide additional numerical evidence of magnetic
reconnection in turbulent environments by testing different mechanisms for
injecting turbulence.  In \S\ref{sec:lv99model} we briefly review the LV99 model
of reconnection and its theoretical predictions.  In \S\ref{sec:setup} we
describe in details the numerical model studied in this paper and new method of
turbulence driving.  Although the initial setup and boundary conditions are
similar to our previous studies, and described in details in \cite{kowal09}, we
briefly describe them here for completeness, as well.  In \S\ref{sec:results} we
present an extensive description of new results obtained from studying our
numeric model, which we discuss later in \S\ref{sec:discussion}.  In
\S\ref{sec:summary} we present our main conclusions.

\section{The Lazarian-Vishniac (1999) Model}
\label{sec:lv99model}

The notion that turbulence can influence reconnection rate is not unprecedented.
 The ideas in this regard were discussed long before LV99.  However, they fell
short of solving the problem.  For instance, \cite{speiser70} considered the
effects of turbulence on microscopic resistivity, \cite{jacobson84} proposed
that the current diffusivity should be modified to include the diffusion of
electrons across the large scale magnetic field due to the small scale field
line stochasticity.  The consequent modifications to the ohmic resistivity have
only a marginal effect on the Sweet-Parker reconnection speeds.
\cite{matthaeus85,matthaeus86} studied 2D magnetic reconnection in the presence
of external turbulence both theoretically and numerically.  They pointed out
various turbulence mechanisms that would enhance reconnection rates, including
multiple X-points as reconnection sites.  However, this work did not include the
effect of magnetic field wandering, which is at the core of the LV99.  They did
not provide analytical predictions of the reconnection speed either\footnote{At
the same time, after LV99 was published, \cite{kim01} produced a study arguing
that turbulence will not change reconnection reconnection rates in the
Sweet-Parker geometry.  This study has been criticized by \cite{lazarian04} and
\cite{eyink11}.  The present paper provides numerical evidence that the
reconnection rates do increase in the presence of turbulence.}.

\begin{figure}
\includegraphics[width=0.45\textwidth]{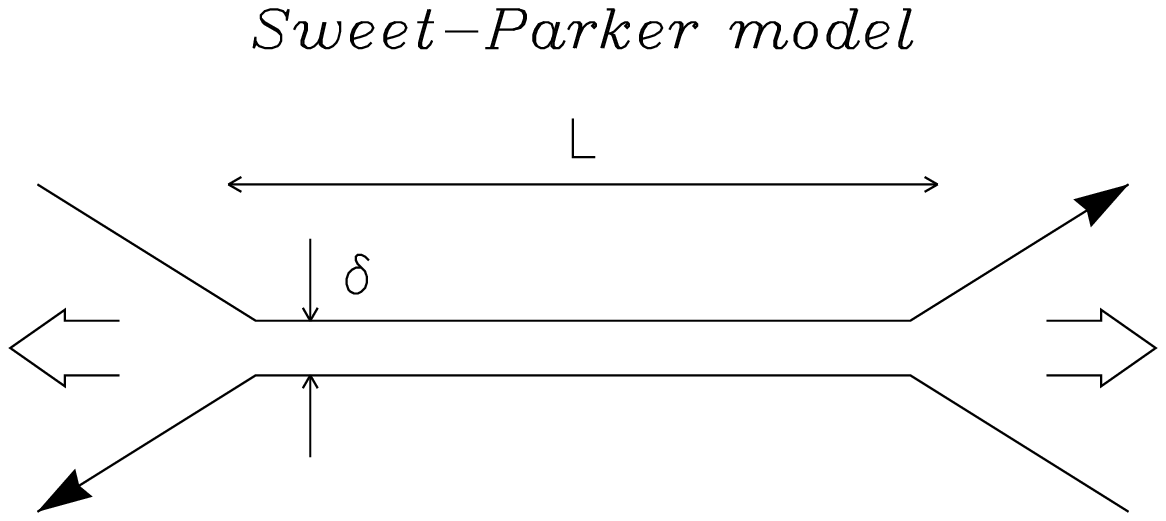}
\includegraphics[width=0.45\textwidth]{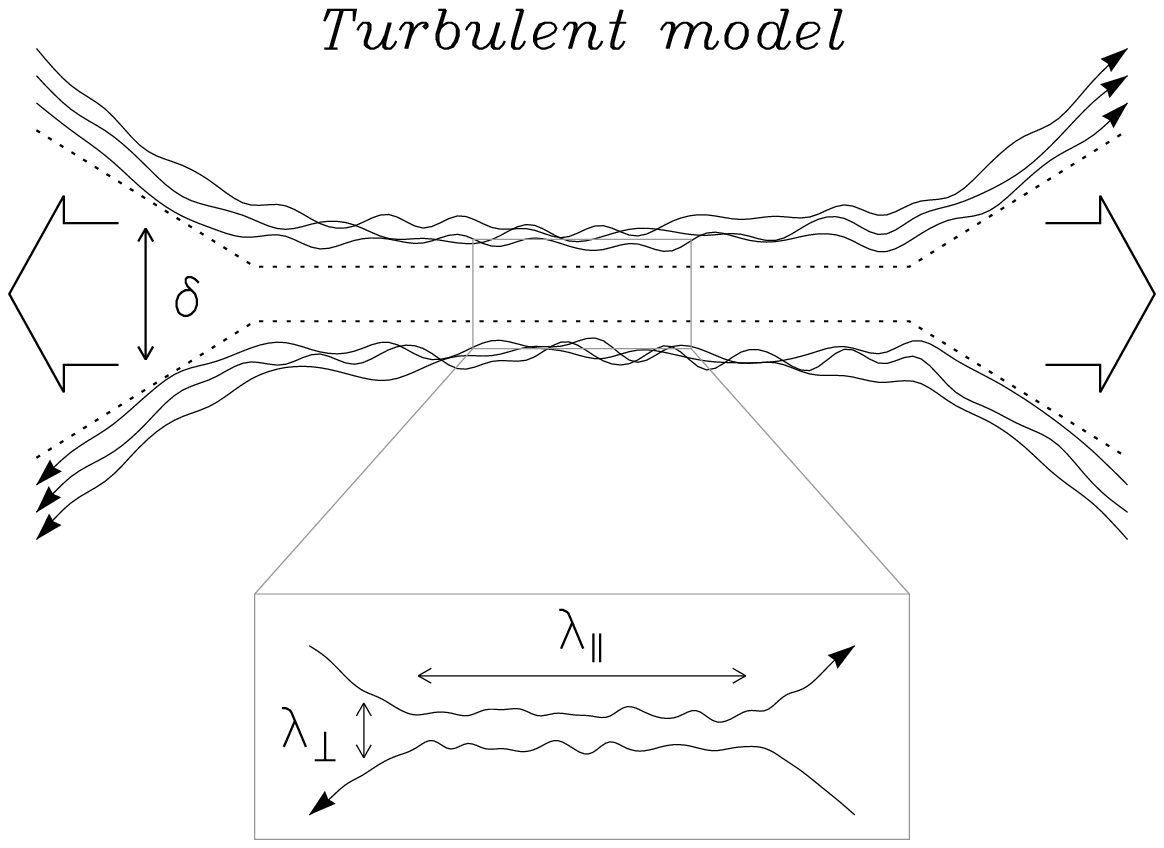}
\caption{{\it Upper plot}: The Sweet-Parker reconnection model.  The outflow is
confined to a thin layer of  $\delta$, which is set by Ohmic diffusivity.  The
length of the current sheet is a macroscopic scale $L \gg \delta$.  Magnetic
field lines are assumed to be laminar.
{\it Middle plot}: Reconnection in the presence of stochastic magnetic field
lines.  The stochasticity introduced by turbulence is weak and the mean field is
clear direction.  The outflow width is set by the diffusion of the magnetic
field lines, which is a macroscopic process, independent of resistivity.
{\it Low plot}: An individual small scale reconnection region.  The reconnection
over small patches of magnetic field determines the local reconnection rate. The
global reconnection rate is substantially larger as many independent patches
reconnect simultaneously.  Conservatively, the LV99 model assumes that the small
scale events happen at a slow Sweet-Parker rate.  Following \cite{lazarian04}
and \cite{kowal09}. \label{fig:lv99model}}
\end{figure}

We begin by offering a brief summary of the differences between the Sweet-Parker
model of the laminar reconnection \citep{parker57,sweet58} and the
Lazarian-Vishniac model which accounts for the effects of turbulence
\citep{lazarian99}.  The latter can be seen as a generalization of the
Sweet-Parker model (see Fig.~\ref{fig:lv99model}) in the sense that the two
regions of differing magnetic directions are pressed up against one another over
a broad contact region.  This is a generic configuration, which should arise
naturally whenever a magnetic field has a non-trivial configuration, whose
energy could be lowered through reconnection.  The outflow of plasma and
reconnected flux will fluctuate as the turbulence evolves and the field line
connections change, but the long term average will reflect the turbulent
diffusion of the field lines.  Consequently, the essential difference between
the Sweet-Parker model and the LV99 model is that the former the outflow is
limited by microphysical Ohmic diffusivity, while in the LV99 model the
large-scale magnetic field wandering determines the thickness of outflow.  For
extremely weak turbulence, when the range of magnetic field wandering becomes
smaller than the width of the Sweet-Parker layer $L S^{-1/2}$, the two models
are indistinguishable.  By weak turbulence, following LV99, we understand a
regime where the correlation length is much greater than the distance by which
individual field lines deviate from a straight line.

LV99 considered a large scale, well-ordered magnetic field, of the kind that is
normally used as a starting point for discussions of reconnection.  In the
presence of turbulence the field has some small scale `wandering'.  LV99
suggested that the presence of a random magnetic field component leads to fast
reconnection.  There are three phenomena mainly responsible for this:
\begin{itemize}
\item only a small fraction of any magnetic field line is subject to direct
Ohmic annihilation, therefore the fraction of magnetic energy that goes directly
into heating the fluid drops down to zero as the fluid resistivity vanishes,
\item the presence of turbulence enables many magnetic field lines to enter the
reconnection zone simultaneously,
\item turbulence broadens the width of the ejection thickness allowing for more
efficient removal of the reconnected flux.
\end{itemize}

With the \citet[][henceforth GS95]{goldreich95} model of turbulence, LV99
obtained:
\begin{equation}
 V_{rec}=V_A \left({l\over L} \right)^{1/2} \left({v_l\over V_A}\right)^{2}, \label{eq:constraint}
\end{equation}
where $l$ and $v_l$ are the energy injection scale and velocity.  This
expression assumes that energy is injected isotropically at the scale $l$
smaller than the length of current sheet $L$, which for subAlfv\'enic turbulence
leads to the generation of weakly interacting waves at that scale.  The waves
transfer energy to modes with larger values of $k_{\perp}$ until strong
turbulence sets in.  It is important to note that the strongly turbulent eddies
have a characteristic velocity of  $v_{turb}\approx V_A(v_l/V_A)^2$.  In other
words, the reconnection speed is the large eddy strong turbulent velocity times
factors which depend on whether the current sheet is smaller or larger than the
large eddies (whose length is approximately the injection scale).  In this sense
the reconnection speed should be fairly insensitive to the exact mechanism for
turbulent power injection.  The main purpose of this paper is test whether or
not this is true for a simple modification of the driving mechanism used in
\citet{kowal09}.

It is important to note three features of Equation~(\ref{eq:constraint}).
First, and most important, it is independent of resistivity.  This is, by
definition, fast reconnection. Second, we usually expect reconnection to be
close to the turbulent eddy speed, the geometric ratios that enter the
expression, i.e. the injection scale $l$ divided by the length of the
reconnection layer $L$, are typically of order unity.  Reconnection will occur
on dynamical time scales. Finally, we note that in particular situations when
turbulence is extremely weak the reconnection speed can be much slower than the
Alfv\'en speed.

More recently, Equation~(\ref{eq:constraint}) was derived using the ideas based
on the well-known concept of Richardson diffusion \citep{eyink11}.  From the
theoretical perspective this new derivation avoids rather complex considerations
of the cascade of reconnection events that were presented in LV99 to justify the
model.  \cite{eyink11} also shows that LV99 model is closely connected to the
recently developed idea of ``spontaneous stochasticity'' of magnetic fields in
turbulent fluids.

In general, the situation in reconnection community now is very different from
that a decade ago.  Currently, possibilities of fast reconnection in MHD regime
due to instabilities of the reconnection layers are widely discussed
\citep{loureiro09,bhattacharjee09}.  These ideas can be traced back to the work
of \cite{shibata01}.  The instabilities, like tearing instability, open up the
reconnection layer enabling a wide outflow.  We expect such an outflow to become
turbulent for most of astrophysical conditions.  In this case, the process can
be important for initiating reconnection in the particular situation when the
level of preexisting turbulence is initially low to initiate sufficiently fast
reconnection.  We feel that exploring the ways of initiation of turbulent
reconnection is very synergistic to the LV99 ideas, but in the current paper we
focus on the case of preexisting turbulence of sufficient level.  This is the
primary domain for which LV99 provides predictions.

Given the limited dynamical range of numerical simulations, we can only inject
power on scales less than than $L$.  The most convenient numerical parameter is
not $v_l$, but the energy injection power $P$.  The power in the turbulent
cascade is $P \sim v_{turb}^2 (V_A/l)$ or $v_l^4/(lV_A)$.  The amount of energy
injected during one Alfv\'en time unit $t_A \equiv L/V_A$, which is constant in
our models, is $t_A P \sim (L/V_A) v_l^4/(lV_A)$. Therefore $v_l^2 \sim
(l/L)^{1/2} (P t_A)^{1/2} V_A$.  Substituting $v_l^2$ in
Equation~(\ref{eq:constraint}) results in
\begin{equation}
 V_{rec} \sim \left( \frac{l}{L} \right) \left( t_A P \right) ^{1/2} \propto l \, P^{1/2},
 \label{eq:scaling}
\end{equation}
which is the prediction we will test here.  In what follows we refer to the
injection power and scale using $P_{inj}$ and $l_{inj}$, respectively.

\section{Numerical Setup}
\label{sec:setup}

\subsection{Governing Equations}
\label{ssec:equations}

We use a high-order shock-capturing Godunov-type scheme based on the
monotonicity preserving (MP) spatial reconstruction
\citep[see][e.g.]{suresh97,he11} and Strong Stability Preserving Runge-Kutta
(SSPRK) time integration \citep[see][and references therein]{gottlieb09} to
solve isothermal non-ideal MHD equations,
\begin{eqnarray}
 \pder{\rho}{t} + \nabla \cdot \left( \rho \vc{v} \right) & = & 0,
 \label{eq:mass} \\
 \pder{\rho \vc{v}}{t} + \nabla \cdot \left[ \rho \vc{v} \vc{v} +
 \left( a^2 \rho + \frac{B^2}{8 \pi} \right) I - \frac{1}{4 \pi} \vc{B} \vc{B}
 \right] & = & \vc{f}, \label{eq:momentum} \\
 \pder{\vc{A}}{t} + \vc{E} & = & \vc{g}, \label{eq:induction}
\end{eqnarray}
where $\rho$ and $\vc{v}$ are plasma density and velocity, respectively,
$\vc{A}$ is the vector potential, $\vc{E} = - \vc{v} \times \vc{B} + \eta \,
\vc{J}$ is the electric field, $\vc{B} \equiv \nabla \times \vc{A}$ is the
magnetic field, $\vc{J} = \nabla \times \vc{B}$ is the current density, $a$ is
the isothermal speed of sound, $\eta$ is the resistivity coefficient, and
$\vc{f}$ and $\vc{g}$ represent the turbulence driving terms either in velocity
or vector potential.  We used a multi-state Harten-Lax-van Leer
\cite[HLLD,][]{mignone07} approximate Riemann solver for solving the isothermal
MHD equations.  The HLLD Riemann solver takes into account magnetic fields, and
can follow Alfv\'en waves with minimal numerical dissipation.  This is
particularly important here, because our simulations are in the
quasi-incompressible regime, where most of energy is transported by Alfv\'en
waves.  The $\nabla \cdot\vec{B} = 0$ is maintained by solving the induction
equation (Eq.~\ref{eq:induction}) using the field interpolated constrained
transport (CT) scheme based on a staggered mesh
\citep[e.g.][]{londrillo00,toth00}.

\subsection{Model Description and Initial Conditions}
\label{ssec:initial}

\begin{figure*}
\center
\includegraphics[width=0.45\textwidth]{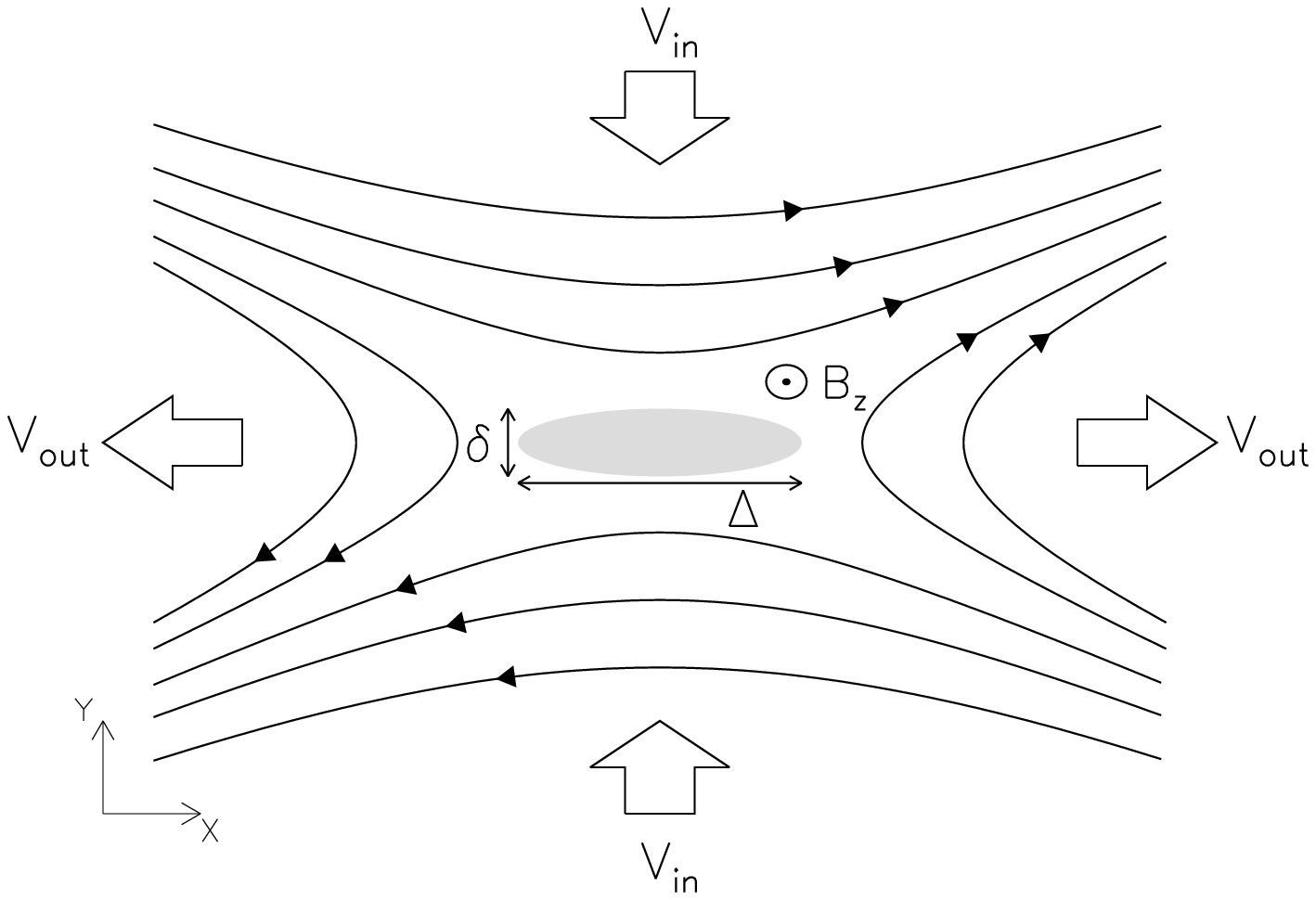}
\includegraphics[width=0.45\textwidth]{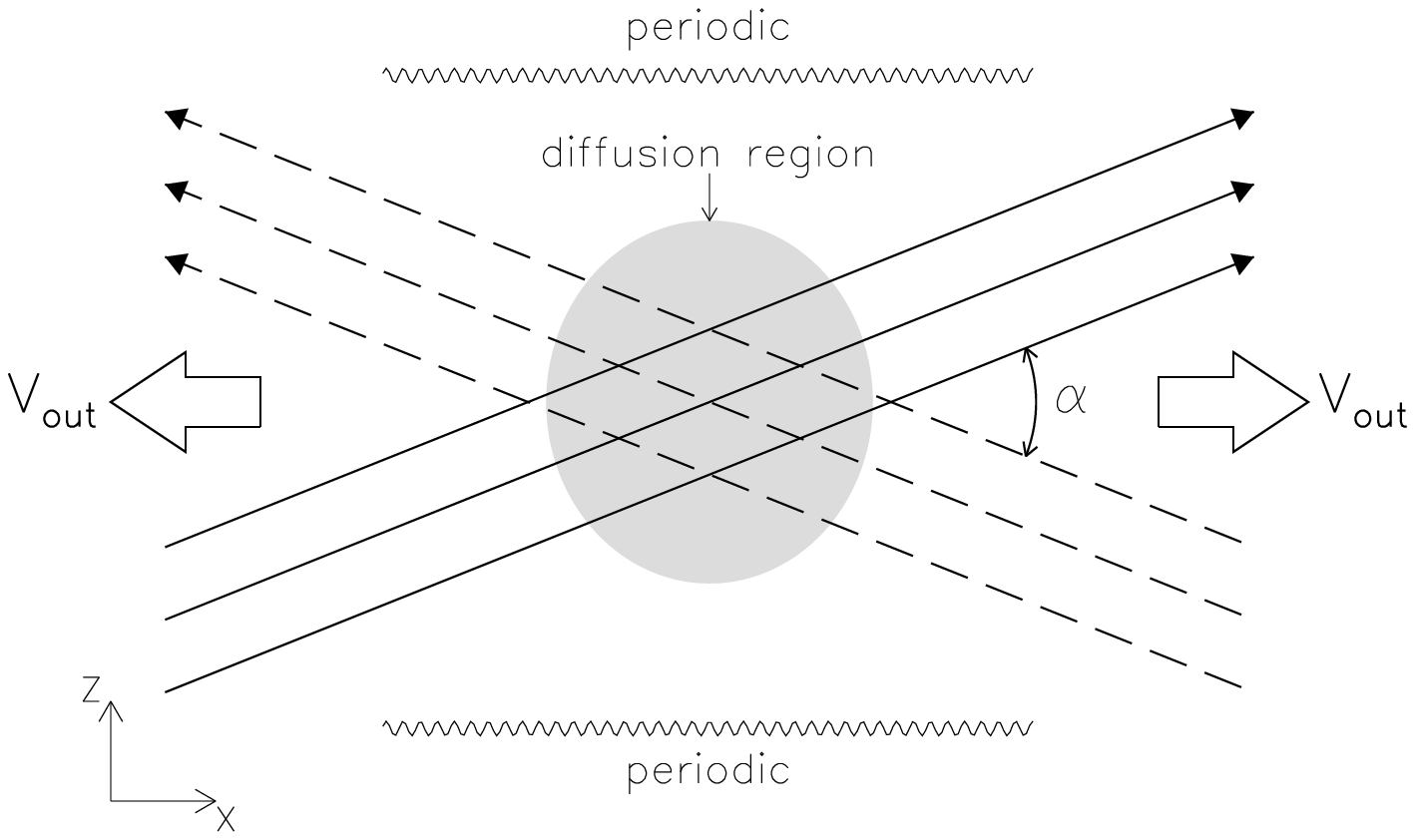}
\caption{A schematic of our magnetic field configuration  projected on the XY
(left) and XZ (right) planes.
{\em Left:} XY projection of the magnetic field lines.  The gray area describes
the diffusion region where the incoming field lines reconnect.  The longitudinal
and transverse scales of the diffusion region aregiven by   $\Delta$ and
$\delta$, respectively.  We use outflow and inflow boundary conditions in the
$\hat x$ and $\hat y$ directions, respectively.
{\em Right:} XZ projection of the magnetic field lines as seen from the top.
Solid and dashed lines show the incoming field lines from the upper and lower
parts of the domain, respectively.  We see that the oppositely directed field
lines are not antiparallel but are set an angle $\alpha$ determined by the
strength of the shared component $B_{z}$. The $\hat z$ boundary conditions can
be  open or periodic, depending on the model. \cite[from][]{kowal09}
\label{fig:setup_planes}}
\end{figure*}

Our reconnection simulation setup is illustrated in
Figure~\ref{fig:setup_planes}, which is a 2D cut through the problem setup,
indicating the location of the diffusion region.  The top and bottom of the
computational domain contain equal and opposite field components in the $\hat x$
direction, as well as a shared component $B_z$ (see the left panel of
Fig.~\ref{fig:setup_planes}).  Magnetic field lines enter through the top and
bottom and are bent by the inflow $V_{in}$ as they move into the diffusion
region.  The diffusion region has a length  $\Delta$ in the $\hat x$ direction
and a thickness $\delta$ in the $\hat y$ direction (see the left panel of
Fig.~\ref{fig:setup_planes}).  The box is periodic in the $\hat z$ direction and
the diffusion region extends through the entire domain.  The projection of the
magnetic topology on the XZ plane shows that the lines in the upper region
(solid lines in the right panel of Fig.~\ref{fig:setup_planes}) and in the lower
region (dashed lines) create an angle $\alpha$ determined by the strength of the
shared component $B_{0z}$.  Once the incoming magnetic lines enter the diffusion
region, they are reconnected and the product of this process is ejected along X
direction with a speed $V_{out}$ (the left panel of
Fig.~\ref{fig:setup_planes}).

We begin with a Harris current sheet of the form $B_x(x,y,z) = B_{0x} \tanh
(y/\theta)$ initialized using the magnetic vector potential $A_z(x,y,z) = \ln |
\cosh(y / \theta) |$,and a uniform guide field $B_z(x,y,z) = B_{0z} =
\mathrm{const}$.  The initial setup is completed by setting the density profile
from the condition of the uniform total (thermal plus magnetic) pressure
$p_{tot}(t=0,x,y,z) = \mathrm{const}$ and setting the initial velocity to zero.

Magnetic reconnection is initiated by a small perturbation of the vector
potential $\delta A_z(x,y,z) = \delta B_{0x} \cos(2 \pi x)$ to the initial
configuration of $A_z(t=0,x,y,z)$ whose strength is given by the coefficient
$\delta B_{0x}$.

In all our simulations the strength of the magnetic field is expressed in terms
of the Alfv\'en velocity defined by the antiparallel component of the
unperturbed magnetic field.  Similarly the density is expressed in terms of the
unperturbed density $\rho_0=1$ and velocities are expressed as fractions of the
fiducial Alfv\'en speed.  The length of the box in the $\hat x$ direction
defines the unit of distance and time is measured in units of $L_x/V_A$.  In the
new set of models we set the initial strength of the antiparallel magnetic field
component $B_{0x}=1.0$ and the guide field $B_{0z}=0.1$.  We performed modeling
for two resistivity coefficients $\eta_u = 5\cdot10^{-4}$ and $\eta_u = 10^{-3}$
which are expressed in the dimensionless units.  The initial perturbation is set
to $\delta B_{0x} = 0.024$.  In order to avoid the complications of strong
compressibility we have set the sound speed to 4.0.

\subsection{Boundary Conditions}
\label{sec:boundaries}

Our computational box has a grid of 256x512x256 or for higher resolution runs,
512x1024x512.  In dimensionless units its size is $L_x=L_z=1$ and $L_y=2$.  We
double the size in the $\hat y$ direction to keep the driven turbulence away
from the inflow boundary.  There is no physical reason to do this, but driving
turbulence near the inflow boundary produces numerical instabilities.

As mentioned earlier, we use three different types of boundary conditions,
depending on the direction of the boundary: outflow boundary conditions along
the $\hat x$ direction, inflow boundary conditions along the $\hat y$ direction
and (sometimes) periodic boundary conditions along the $\hat z$ direction.

The open boundary conditions are the same as those used in our previous
modeling.  We refer to \cite{kowal09} for their detailed description.  Briefly,
we use simple ``zero-gradient'' boundary conditions, setting the normal
derivatives of the fluid variables (density and momentum) to zero.  In the
hydrodynamic limit this allows waves to leave the box without significant
boundary reflections.  In turbulence simulations this can lead to a slight drift
in the fluid density.  There is no requirement that the boundary density is
constant, and inflows and outflows can cause a small net gain or loss from the
system.  Fortunately, changes in the total mass are small and only fluctuate
around the initial value \citep{kowal09}.  They do not influence our results
significantly.

In order to incorporate the magnetic field into the open boundary conditions, we
set the transverse components of the vector potential $\vc{A}$ using first order
extrapolation.  The normal derivative of the normal component is set to zero.
In this way the normal derivatives of the transverse components of the magnetic
field are zero, while the normal component of magnetic field is calculated from
the zero-divergence condition $\nabla \cdot \vc{B} = 0$.  This approach avoids
the generation of nonzero magnetic divergence at the boundary.  However, it has
the drawback that it creates a small jump in the momentum flux across the
boundary resulting from the presence of non-zero terms $\left( - B_x, B_y, B_z
\right) \partial_x B_x$ at the X outflow boundary and $\left( B_x, - B_y, B_z
\right) \partial_y B_y$ at the Y inflow boundary.  We have evaluated the
velocity increment these terms produce at each time step.  In models with the
strongest turbulence these terms were of order of $10^{-6}$ and $10^{-8}$ at the
X and Y boundaries, respectively.  In the presence of strong outflows and
inflows, generally of order unity, they are clearly negligible.

Simulations with explicit resistivity run into problems at the boundaries.  In
order to avoid a non-continuous resistive term and difficulties with the
treatment of the current density $\vc{J}$ we have introduced a zone of decaying
resistivity near the boundary.  In a thin layer near the boundary, the value of
resistivity $\eta_u$ decays down to a very small value chosen to be close to the
numerical resistivity $\eta_n$ of our code.  In our models we adopt the value of
$\eta_n = 3 \cdot 10^{-4}$.  None of this has an effect on  the reconnection
speeds.  The validation of this method was presented in \cite{kowal09}.

\subsection{New Method of Turbulence Driving}
\label{ssec:forcing}

In our previous work we drove turbulence using a method described by
\cite{alvelius99}, in which the driving term was implemented in the spectral
space with discrete Fourier components concentrated around a wave vector
$k_{inj}$ corresponding to the injection scale $l_{inj} = 1/k_{inj}$.  We
perturbed a number $N_f$ of discrete Fourier components of velocity in a shell
extending from $k_{inj}-\Delta k_{inj}$ to $k_{inj}+\Delta k_{inj}$ with a
Gaussian profile of the half width $k_c$ and the peak amplitude $\tilde{v}_f$ at
the injection scale.  The amplitude of driving is solely determined by its power
$P_{inj}$, the number of driven Fourier components and the time step of driving
$\Delta t_f$.  The randomness in time makes the force neutral in the sense that
it does not directly correlate with any of the time scales of the turbulent
flow, and it also determines the power input solely by the force-force
correlation.

On the right hand side of Equation~(\ref{eq:momentum}), the forcing is
represented by a function $\vc{f} = \rho \dot{\vc{u}}$, where $\rho$ is local
density and $\dot{\vc{u}}$ is random acceleration calculated using the method
described above.  In a similar way we can drive turbulence in the vector
potential or magnetic field, which is represented by term $\vc{g}$ on the right
hand side of the induction equation (Eq.~\ref{eq:induction}).

In the new method of turbulence driving we add individual eddies with random
locations of their centers and random orientations, either to velocity or
magnetic field, at random moments in time.  This guarantees the randomness of
new forcing.

Each eddy is calculated from a kernel function described by a directional vector
$\vc{a}$ (with amplitude $|\vc{a}|$) multiplied by a Gaussian function
\begin{equation}
 \mathbf{\Psi}(\vc{r}) = \vc{a} \exp\left( - \frac{\left|  \vc{r} - \vc{r}_c\right|^2}{2 \delta^2} \right),
\end{equation}
where $\vc{r}_c$ is the location of the eddy center, and $\delta$ is the eddy
width.  An actual eddy is generated from such a kernel function by taking its
curl, i.e. $\delta \vc{f} = \rho \left( \nabla \times \mathbf{\Psi} \right) d t$
or $\delta \vc{g} = \nabla \times \mathbf{\Psi} d t$ in the case of injection in
velocity or magnetic field, respectively.  For example, if we assume that we
inject one eddy in the magnetic field at $\vc{r}_c = \left( 0, 0, 0 \right)$,
and that the perturbing vector potential fluctuation has only the nonzero
component, i.e. $\mathbf{\Phi} = (0, 0, \Phi_z)$, the contribution to magnetic
field is expressed by
\begin{equation}
 \left[
 \begin{array}{c}
  \delta g_x \\
  \delta g_y \\
  \delta g_z
 \end{array}
 \right] \left( x, y, z \right) = \frac{|\vc{a}|}{\delta^2} \exp\left( - \frac{|\vc{r}|^2}{2 \delta^2} \right)
 \left[
 \begin{array}{c}
  - y \\
    x \\
    0
 \end{array}
 \right] dt \ .
\end{equation}
This function describes an eddy injected in the XY plane with the maximum
amplitude $\delta g_{max} = |\vc{a}| \delta^{-1} e^{-\frac{1}{2}}$ at the
distance $r_{max} = \delta$ and injection scale $l_{inj} = \delta$.  We
know the energy injected by one eddy, which is $\Delta E_{eddy} = \pi^{3/2}
|\vc{a}|^2 \delta / 2$, therefore we can determine its amplitude $|\vc{a}|$ from
the injection power $P_{inj}$ and the injection rate $N_{inj}$, which is the
number of eddies injected in a time unit,
\begin{equation}
 P_{inj} = N_{inj} \Delta E_{eddy} \ \rightarrow \  |\vc{a}| = \sqrt{\frac{2 P_{inj}}{\pi^{3/2} N_{inj} \delta}} \ .
\end{equation}
These estimates are done for the 3D case.  In the 2D case the eddy energy is
$\Delta E_{eddy} = \pi |\vc{a}|^2 / 2$, and therefore the eddy amplitude can be
determined from $|\vc{a}| = \sqrt{2 P_{inj} / (\pi N_{inj})}$.

In the new method there is no direct treatment of the velocity-force
correlation, therefore there is no guarantee that this correlation is zero and
the injected power is completely determined by the force-force correlation.  A
reasonable solution to this problem would be to control the amount of injected
energy and modify the amplitude of injected eddies or the injection rate at each
time step in order to compensate the differences in the energy injected in the
domain.  The performed tests show, however, that although the velocity-force
correlation is not zero, it is in fact fluctuating in time with a small
amplitude, and giving in result a zero net contribution.

The new method drives turbulence directly in the real space, in contrast to the
previous one, therefore it can be applied locally.  We drive turbulence in a
subvolume of the domain.  The size of the subvolume is determined by two scales,
the radius $r_d$ on the XZ plane around the center of the domain and the height
$h_d$ describing the thickness of the driving region from the midplane.  In this
way we avoid driving turbulence at the boundary and reduce the influence of
driving on the inflow or outflow.

All models are evolved without turbulence for several dynamical times in order
to allow the system to achieve stationary laminar reconnection.  Then, at a
given time $t_b$ we start driving turbulence by increasing its amplitude to the
desired level, until $t_e$.  In this way we let the system to adjust to a new
state.  From time $t_e$ the turbulence is driven with the full power $P_{inj}$.

\subsection{Reconnection Rate Measure}
\label{sec:rec_rate}

In the next sections we measure the reconnection rate using the new method of
reconnection rate measure introduced in \cite{kowal09} and described by a
formula
\begin{equation}
V_\mathrm{rec} = \frac{1}{2 |B_{x,\infty}| L_z} \left[ \oint{\mathrm{sign} (B_x) \vc{E} \cdot d \vc{l}} - \partial_t \int {|B_x| dA} \right]
\end{equation}
where $B_x$ is the strength of reconnecting magnetic component, $\vc{E}$ is the
electric field, $d\vc{A}$ is area element of an XZ plane across which we perform
integration, $d\vc{l}$ is the line element separating two regions of the YZ
plane defined by the sign of $B_x$, $|B_{x,\infty}|$ is the asymptotic absolute
value of $B_x$, and $L_z$ is the width of the box.

This method of the reconnection rate measure was derived from the magnetic flux
conservation $\Psi$ and takes into account all processes contributing to the
change of magnetic flux.  The electric field $\vc{v} \times \vc{B} - \eta
\vc{j}$ can be further divided into an advection term $\vc{v} \times B_x
\hat{x}$, a shear term $\vc{v} \times \left( B_y \hat{y} + B_z \hat{z} \right)$,
and a resistive term $- \eta \vc{j}$. With this in mind the line integral can be
rewritten as
\begin{eqnarray}
\oint{\mathrm{sign}\left( B_x \right) \vc{E} \cdot d \vc{l}} = \oint{|B_x| \left( \vc{v}_\perp \times \hat{x} \right) \cdot d \vc{l}} & & \\
+ \oint{\mathrm{sign} \left( B_x \right) v_x \left( \hat{x} \times \vc{B}_\perp \right) \cdot d \vc{l}} & - & \oint{ \eta \vc{j} \cdot d \vc{l}} . \nonumber
\end{eqnarray}

This new reconnection measure contains the time derivative of the absolute value
of $B_x$, and a number of boundary terms, such as advection of $B_x$ across the
boundary and the boundary integral of the resistive term $\eta \vc{j}$.  The
additional terms include all processes contributing the time change of $|B_x|$.
In particular, they can have non-zero values.

\subsection{Table of Simulated Models}
\label{sec:models}

\begin{table*}[t]
 \caption{List of models. \label{tab:models}}
 \vskip4mm
 \centering
 \begin{tabular}{c|cccccccccccc}
  \tophline
  Name & $B_{0z}$ & $\eta_u$ [$10^{-3}$] & $\nu_u$ [$10^{-3}$] & $P_{inj}$ & $\Delta k_{inj}$ & Driving Type \\
  \middlehline
PD & 0.1 & 1.0 & 0.0 & 0.1 &  8 & old in $\vc{V}$ \\
{} & 0.1 & 1.0 & 0.0 & 0.2 &  8 & old in $\vc{V}$ \\
{} & 0.1 & 1.0 & 0.0 & 0.5 &  8 & old in $\vc{V}$ \\
{} & 0.1 & 1.0 & 0.0 & 1.0 &  8 & old in $\vc{V}$ \\
{} & 0.1 & 1.0 & 0.0 & 2.0 &  8 & old in $\vc{V}$ \\
{} & 1.0 & 1.0 & 0.0 & 0.1 &  8 & old in $\vc{V}$ \\
{} & 1.0 & 1.0 & 0.0 & 0.2 &  8 & old in $\vc{V}$ \\
{} & 1.0 & 1.0 & 0.0 & 0.5 &  8 & old in $\vc{V}$ \\
{} & 1.0 & 1.0 & 0.0 & 1.0 &  8 & old in $\vc{V}$ \\
{} & 1.0 & 1.0 & 0.0 & 2.0 &  8 & old in $\vc{V}$ \\
$\rightarrow$ & 0.1 & 1.0 & 0.0 & 0.2 &  8 & new in $\vc{B}$ \\
$\rightarrow$ & 0.1 & 1.0 & 0.0 & 0.5 &  8 & new in $\vc{B}$ \\
$\rightarrow$ & 0.1 & 1.0 & 0.0 & 1.0 &  8 & new in $\vc{B}$ \\
HR & 0.1 & 0.5 & 0.0 & 1.0 &  8 & new in $\vc{V}$ \\
\hline
SD & 0.1 & 1.0 & 0.0 & 1.0 &  5 & old in $\vc{V}$ \\
{} & 0.1 & 1.0 & 0.0 & 1.0 &  8 & old in $\vc{V}$ \\
{} & 0.1 & 1.0 & 0.0 & 1.0 & 12 & old in $\vc{V}$ \\
{} & 0.1 & 1.0 & 0.0 & 1.0 & 16 & old in $\vc{V}$ \\
{} & 0.1 & 1.0 & 0.0 & 1.0 & 25 & old in $\vc{V}$ \\
{} & 1.0 & 1.0 & 0.0 & 1.0 &  5 & old in $\vc{V}$ \\
{} & 1.0 & 1.0 & 0.0 & 1.0 &  8 & old in $\vc{V}$ \\
{} & 1.0 & 1.0 & 0.0 & 1.0 & 12 & old in $\vc{V}$ \\
{} & 1.0 & 1.0 & 0.0 & 1.0 & 16 & old in $\vc{V}$ \\
{} & 1.0 & 1.0 & 0.0 & 1.0 & 25 & old in $\vc{V}$ \\
$\rightarrow$ & 0.1 & 1.0 & 0.0 & 1.0 &  8 & new in $\vc{B}$ \\
$\rightarrow$ & 0.1 & 1.0 & 0.0 & 1.0 & 24 & new in $\vc{B}$ \\
$\rightarrow$ & 0.1 & 1.0 & 0.0 & 1.0 & 32 & new in $\vc{B}$ \\
HR & 0.1 & 0.5 & 0.0 & 1.0 &  8 & new in $\vc{V}$ \\
\hline
VD & 0.1 & 1.0 & 0.2 & 1.0 &  8 & old in $\vc{V}$ \\
{} & 0.1 & 1.0 & 0.5 & 1.0 &  8 & old in $\vc{V}$ \\
{} & 0.1 & 1.0 & 1.0 & 1.0 &  8 & old in $\vc{V}$ \\
{} & 0.1 & 1.0 & 2.0 & 1.0 &  8 & old in $\vc{V}$ \\
{} & 0.1 & 1.0 & 3.0 & 1.0 &  8 & old in $\vc{V}$ \\
{} & 0.1 & 1.0 & 4.0 & 1.0 &  8 & old in $\vc{V}$ \\
{} & 0.1 & 1.0 & 5.0 & 1.0 &  8 & old in $\vc{V}$ \\
 \bottomhline
 \end{tabular}
\end{table*}

In Table~\ref{tab:models} we list parameters of all the models presented in this
paper including models from \cite{kowal09} and models with new driving.  As in
the previous paper we divided them into several groups.  In each group we
calculated models in order to study the dependence of the reconnection rate on a
characteristic parameter of turbulence or resistivity.  We have studied the
dependence of reconnection on the power of turbulence (models ''PD''), injection
scale (models ''SD''), and viscosity (models ''VD'').  Models with new driving
are marker with a right arrow ($\rightarrow$), and models with new driving and
higher resolution are marked with a symbol ''HR''.

Only the varying parameters are listed in the table, the strength of guide field
$B_{0z}$, the uniform resistivity $\eta_u$, the uniform viscosity $\nu_u$, the
power of turbulence $P_{inj}$ and its injection scale $k_{inj}$, and the method
of turbulence driving.

All models presented in this section were calculated with the grid size $\Delta
x \approx 0.004$ corresponding to the resolution 256x512x256, except the model
marked with symbol ''HR'', which was simulated with the resolution 512x1024x512
($\Delta x \approx 0.002$).

\section{Results}
\label{sec:results}

\subsection{Time Evolution of Energies and Reconnection Rate}
\label{ssec:evolution}

\begin{figure}
\center
\includegraphics[width=0.45\textwidth]{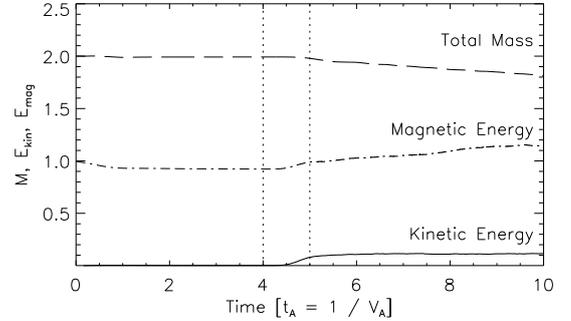}
\caption{Evolution of total mass $M$, and kinetic and magnetic energies,
$E_{kin}$ and $E_{mag}$, respectively.  Two dotted vertical lines bound the
period of gradually increasing turbulence.  The resistivity in this model is set
to $\eta=10^{-3}$ and the shared component of magnetic field $B_{0z} = 0.1$.  In
this model we inject turbulence in the magnetic field. \label{fig:energies}}
\end{figure}

In Figure~\ref{fig:energies} we present an example of the evolution of total
mass, and kinetic and magnetic energies in a model with $P_{in} = 1.0$,
$k_{f}=8$, and $\eta_u=10^{-3}$.  We inject turbulence in magnetic field using
new forcing method, gradually increasing its strength from $t=4$ to $t=5$.  This
period is marked by two dotted vertical lines in Figure~\ref{fig:energies}.  We
see an increase of kinetic energy during this period due to the injection and
saturation after $t=5$.  The kinetic energy preserves constant value during the
turbulent stage very well.  The magnetic energy increases during this stage,
slowly saturating.  This increase is attributed to the injection of magnetic
eddies.  On the contrary, the total mass in the system decays slowly. We
emphasize, that since we use open boundary conditions, not perfect conservation
of mass and energies is possible in the presence of turbulence.

\begin{figure}
\center
\includegraphics[width=0.45\textwidth]{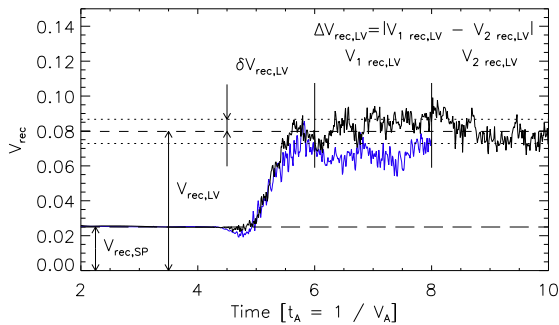}
\caption{Evolution of the reconnection rate $V_{rec}$ (black) for the same model
as in Fig.~\ref{fig:energies}.  Blue line show the evolution of reconnection in
a model with the same parameters in which the turbulence were driven using the
old method.  In this plot we present the measured rates of the Sweet-Parker
reconnection $V_{rec,SP}$ and during the presence of turbulence, $V_{rec,LV}$.
Symbol $\delta V_{rec,LV}$ is the time variance.  $\Delta V_{rec,LV}$ is the
estimated uncertainty of the measure. \label{fig:rate}}
\end{figure}

In Figure~\ref{fig:rate} we show the evolution of reconnection rates $V_{rec}$
for two models with the same set of initial conditions, but in the first model
we drove turbulence by injecting magnetic eddies using the new method described
in this paper (black line), and in the second model we inject velocity
fluctuations using the old method described in \cite{kowal09} (blue line).  In
this plot we recognize an increase of both rates during the introduction of
turbulence.  After the transition period between $t=4$ and $t=5$, when the
system adjusts to a new state, both measures coincide and even though they
fluctuate, they reach a stationary state characterized by faster reconnection.
Both types of turbulence bring the reconnection rate to similar level.  Somewhat
higher reconnection rate in the model with new driving could be attributed to
the fact that this model was calculated using the 5th order spatial
reconstruction and the 3rd order integration in time, in contrast to the old
model where we used the second order methods.  Lower order, especially in the
spatial interpolation, introduces additional numerical diffusion decreasing the
amplitudes of turbulent fluctuations at scales comparable to the current sheet
scale.

In Figure~\ref{fig:rate} we also show the way of measuring the reconnection
rates, in the Sweet-Parker and LV99 stages, $V_{rec,SP}$ and $V_{rec,LV}$,
respectively.  Because the reconnection rates fluctuate in the presence of
turbulence we also measure their time variance $\delta V_{rec,LV}$ using the
standard deviation.  In addition to the time variance of $V_{rec}$, we measure
their errors by splitting the averaging region into two subregions and after
averaging the rates $V_{1 rec}$ and $V_{2 rec}$ over each subregion (see
Fig.~\ref{fig:rate}), we take the absolute value of their difference $\Delta
V_{rec} = V_{1 rec} - V_{2 rec}$.  This difference corresponds to the error of
$V_{rec}$, i.e. it is different from zero if the rate is not constant in time.
In all further analysis and presented plots we use values estimated in this way.
 These measures correspond exactly to those presented in \cite{kowal09}.

\subsection{Topology of Magnetic Field}
\label{ssec:topology}

In this section we compare field topologies in two example models run with the
same set of parameters, but with different types of driving.  Both models have
been simulated with the uniform resistivity $\eta_u=10^{-3}$ and the resolution
256x512x256.  We injected turbulence with power $P_{inj} = 1.0$ at the injection
scale $k_{inj}=8$.  The only difference between models is the way we injected
turbulence.  In the old model we inject velocity fluctuations with random phases
in Fourier space, and then transform them to real space and shape by a window in
order to limit the injection to the specified region near the current sheet.  In
the new model, we inject magnetic loops with random locations and random
orientations in the 3D volume near the current sheet.  The way of injecting
turbulence is essentially different in both cases.

\begin{figure*}
\center
\includegraphics[width=0.33\textwidth]{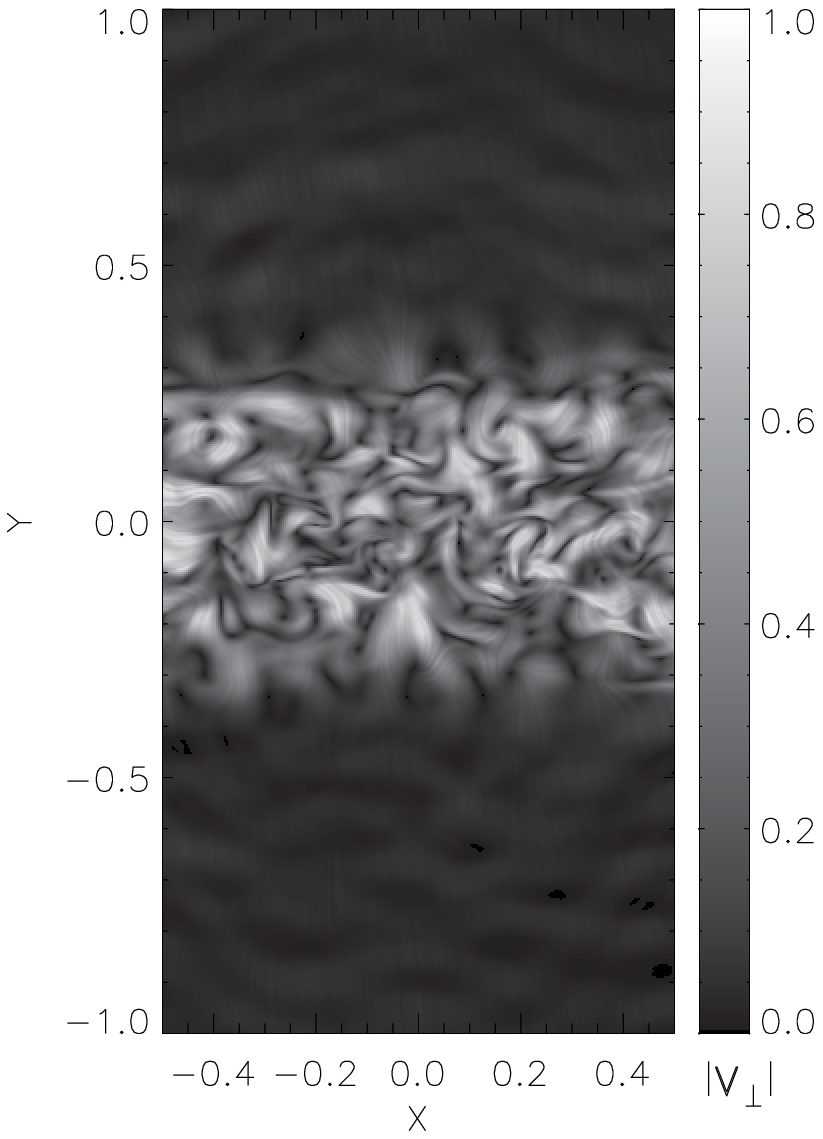}
\includegraphics[width=0.33\textwidth]{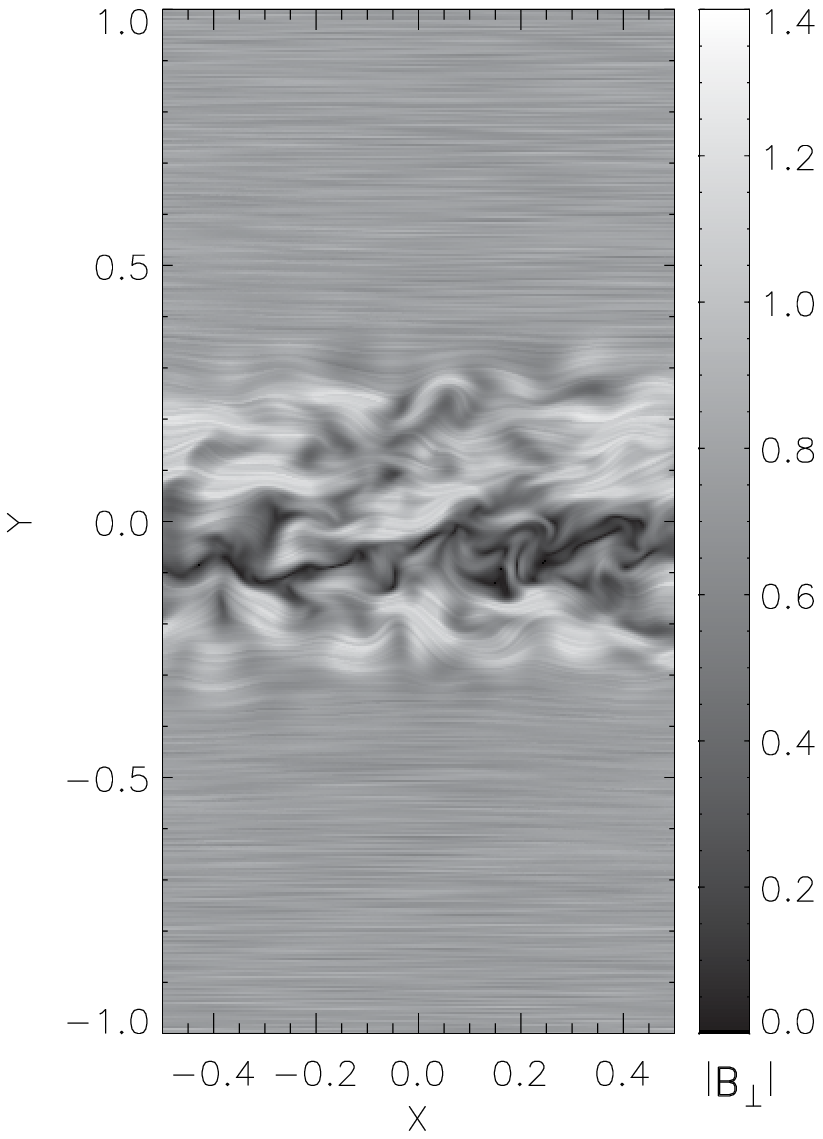}
\includegraphics[width=0.33\textwidth]{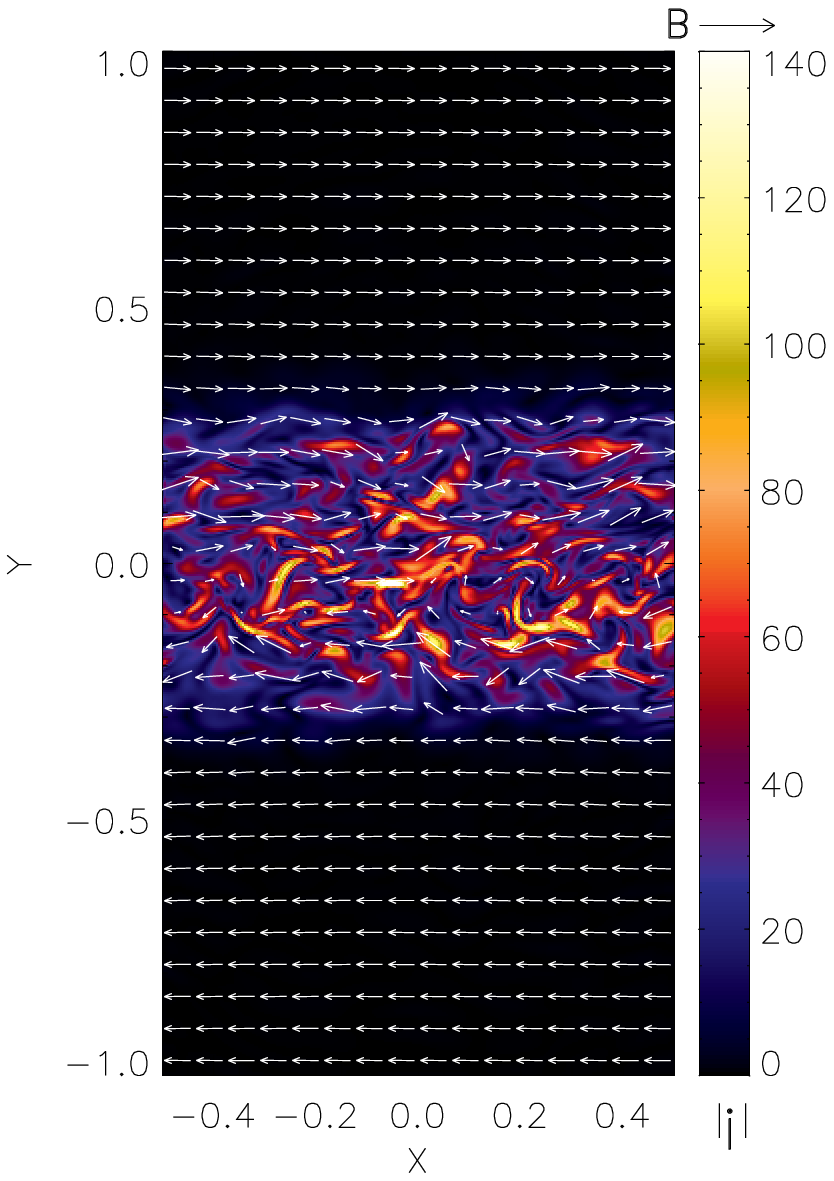}
\includegraphics[width=0.33\textwidth]{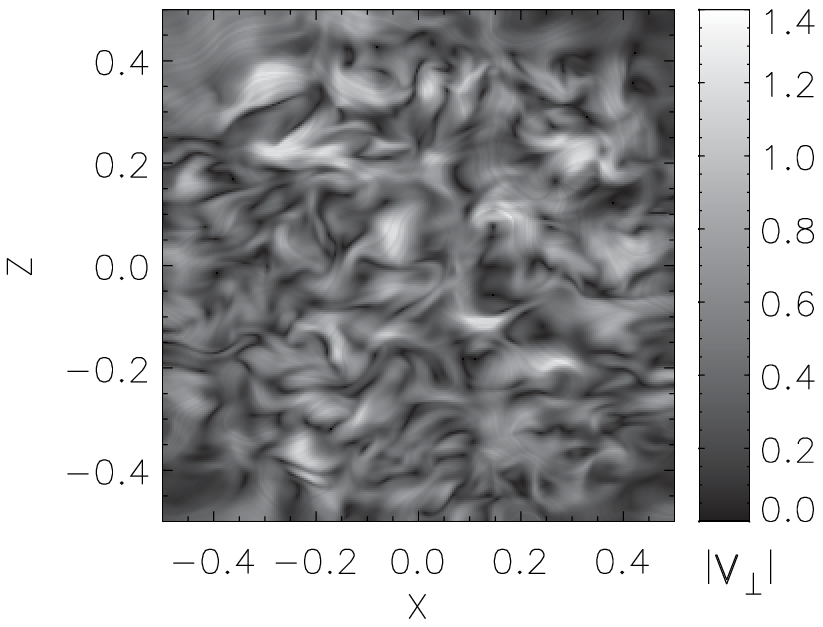}
\includegraphics[width=0.33\textwidth]{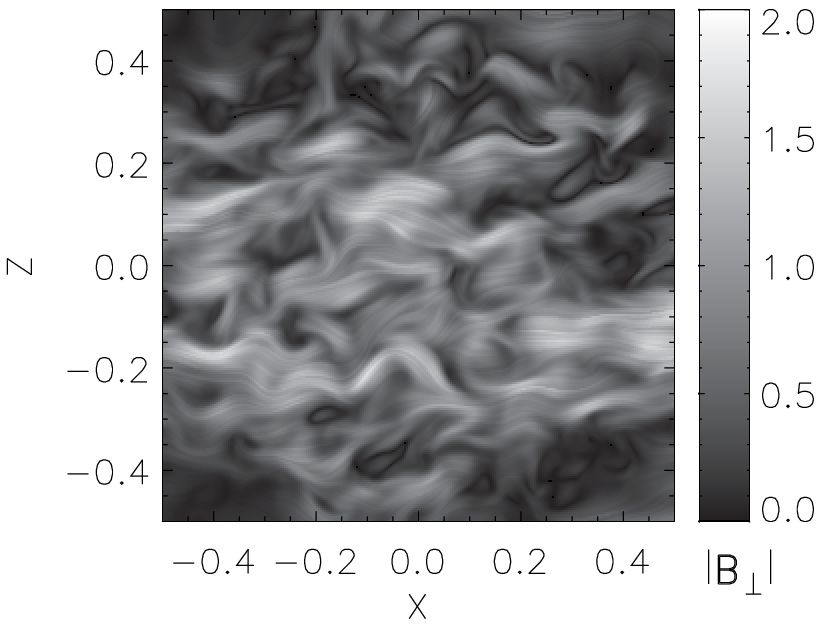}
\includegraphics[width=0.33\textwidth]{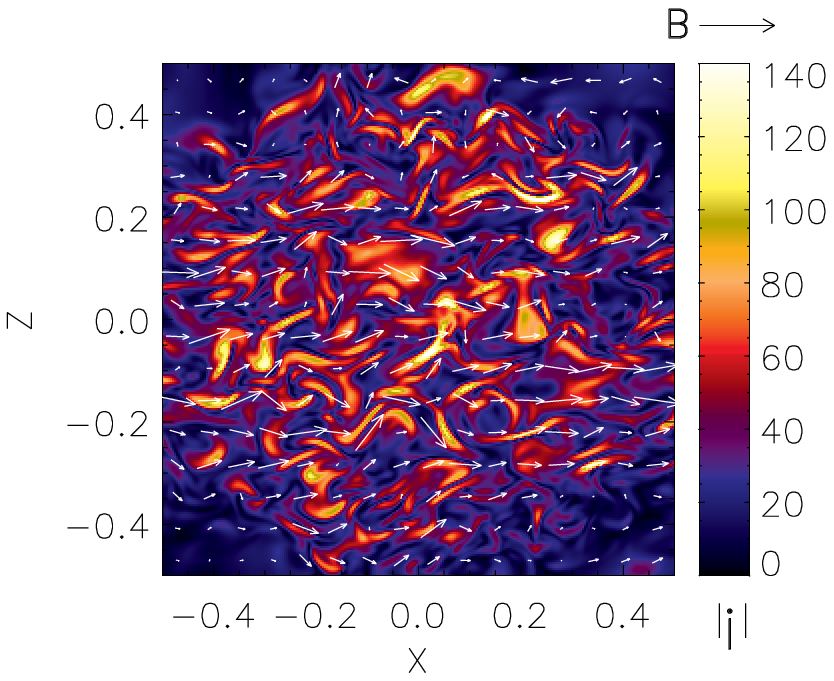}
\caption{Topology and strength of the velocity field (left panel) and magnetic
field (middle panel) in the presence of fully developed turbulence for an
example model with old driving at time $t=12$.  In the right panel we show
distribution of the absolute value of current density $|\vec{J}|$ overlapped
with the magnetic vectors.  The images show the XY-cut (upper row) and XZ-cut
(lower row) of the domain at the midplane of the computational box.  Turbulence
is injected with power $P_{inj}=1$ at scale $k_{inj}=8$ into velocity.  Magnetic
field reversals observed are due to magnetic reconnection rather than driving of
turbulence, which is subAlfv\'enic. \label{fig:top_turb}}
\end{figure*}

In Figure~\ref{fig:top_turb} we show examples of XY-cuts (upper row) and XZ-cuts
(lower row) through the box for the model with old driving.  In the left and
middle columns we show topologies of the velocity and  magnetic field,
respectively, with the intensities corresponding to the amplitude of components
parallel to the plotted plane.  In the right column we show the absolute value
of current density with overplotted magnetic vectors.  Velocity has a very
complex and mixed structure near the midplane due to constant injection of
fluctuations in this region (see the left panel in Fig.~\ref{fig:top_turb}).
The majority of the velocity fluctuations is perpendicular to the mean magnetic
field.  This is because we are in the nearly incompressible regime of turbulence
(large plasma $\beta$) and most of the fluctuations propagate as Alfv\'en waves
along the mean magnetic field.  Slow and fast waves, whose strengths are
significantly reduced, are allowed to propagate in directions perpendicular to
the mean field as well.  As a result, a big fraction of the turbulent kinetic
energy leaves the box along magnetic lines.  We observe, however, an efficient
bending of magnetic lines at the midplane where the field is weaker (see the
upper middle plot in Fig.~\ref{fig:top_turb}).  This is not result of a driving,
but result of reconnection.  In general the interface between positively and
negatively directed magnetic lines is much more complex than in the case of
Sweet-Parker reconnection.  This complexity favors creation of enhanced current
density regions, where the local reconnection works faster (see the right panel
of Fig.~\ref{fig:top_turb}).  Since we observe multiple reconnection events
happening at the same time, the global reconnection rate should be significantly
enhanced.

\begin{figure*}
\center
\includegraphics[width=0.33\textwidth]{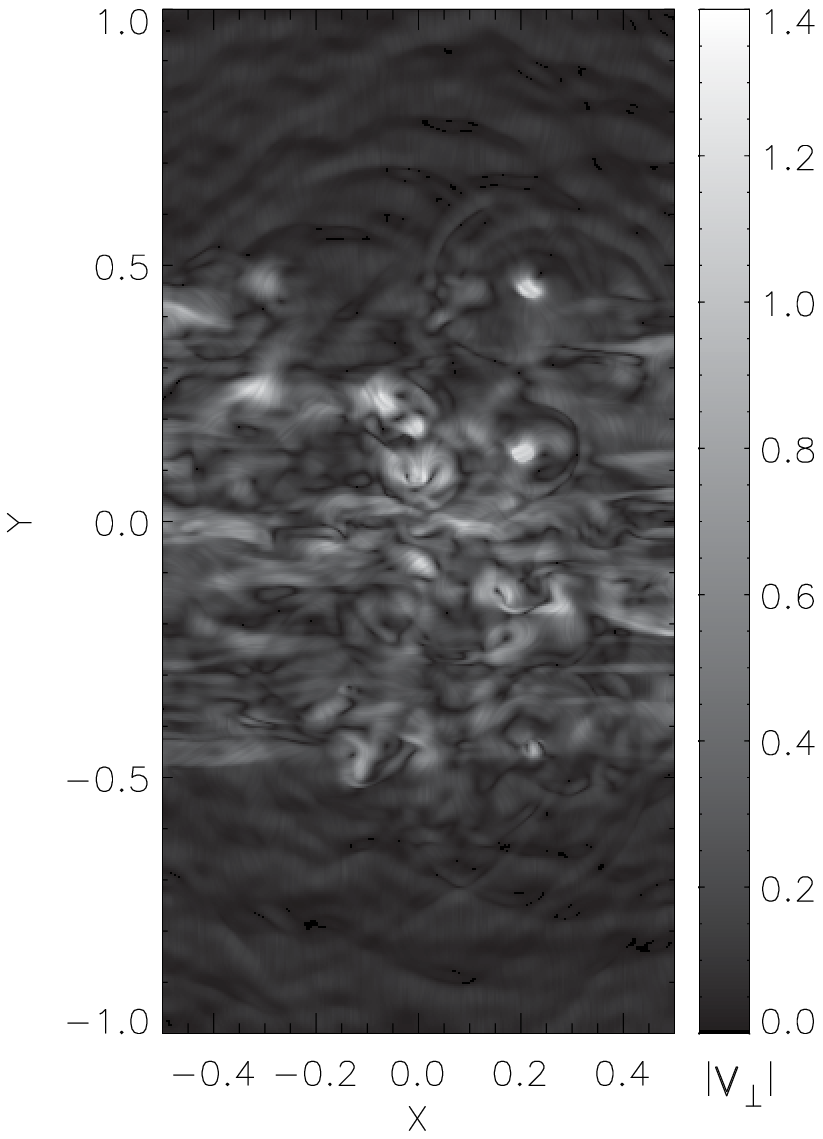}
\includegraphics[width=0.33\textwidth]{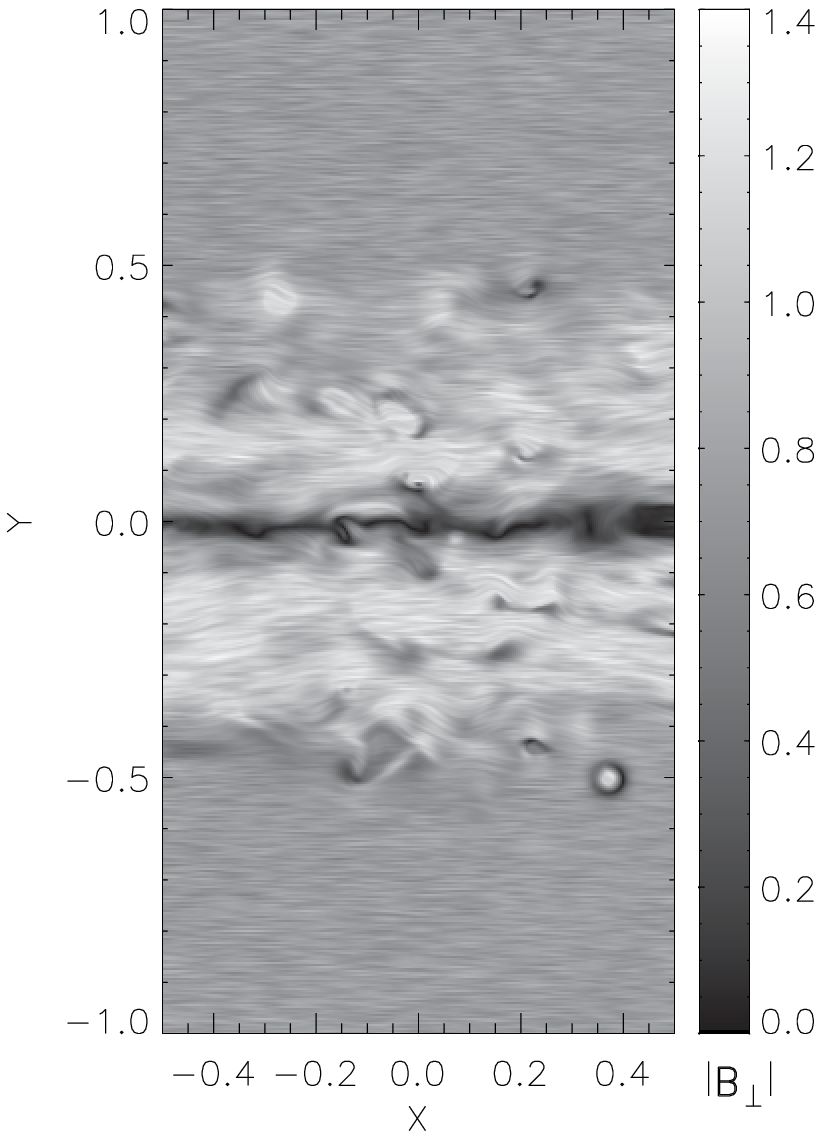}
\includegraphics[width=0.33\textwidth]{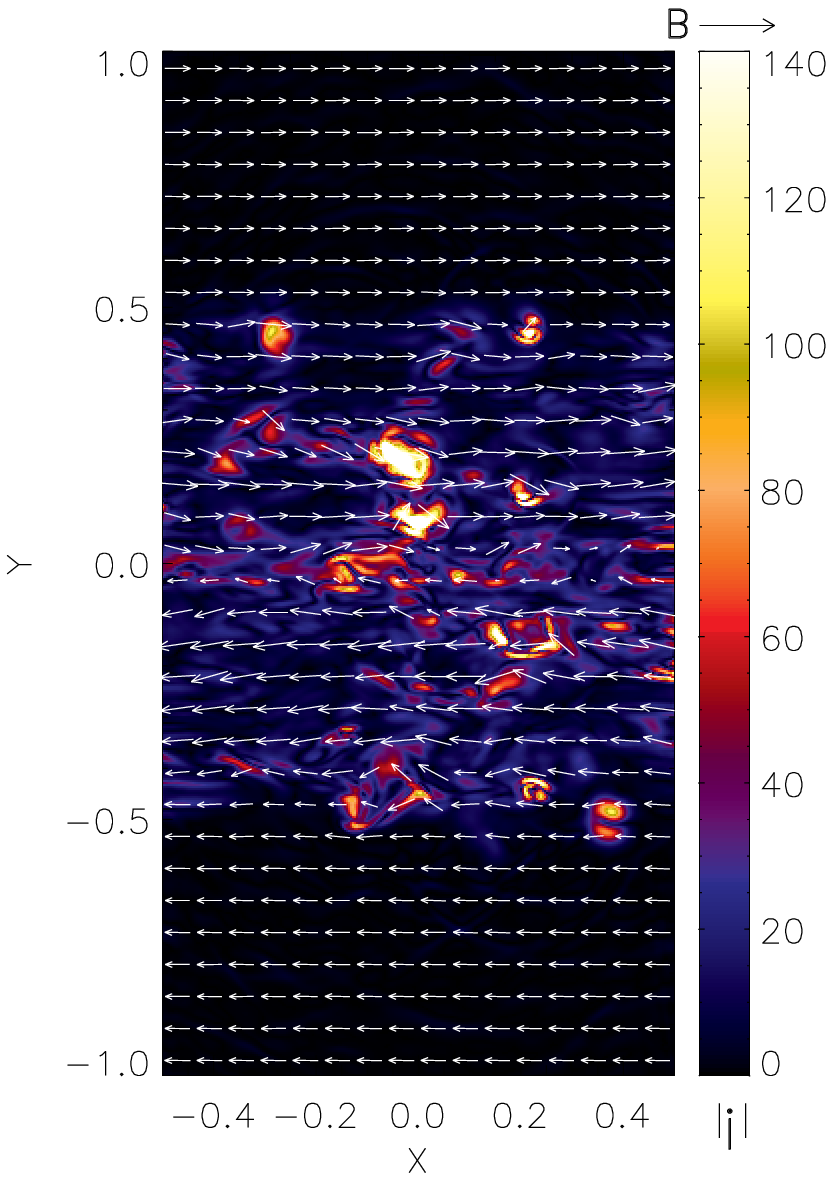}
\includegraphics[width=0.33\textwidth]{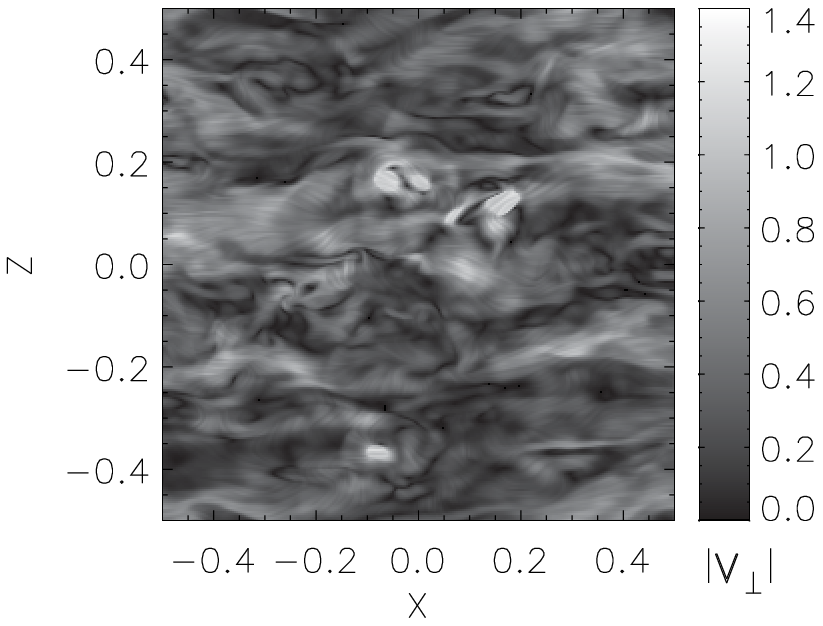}
\includegraphics[width=0.33\textwidth]{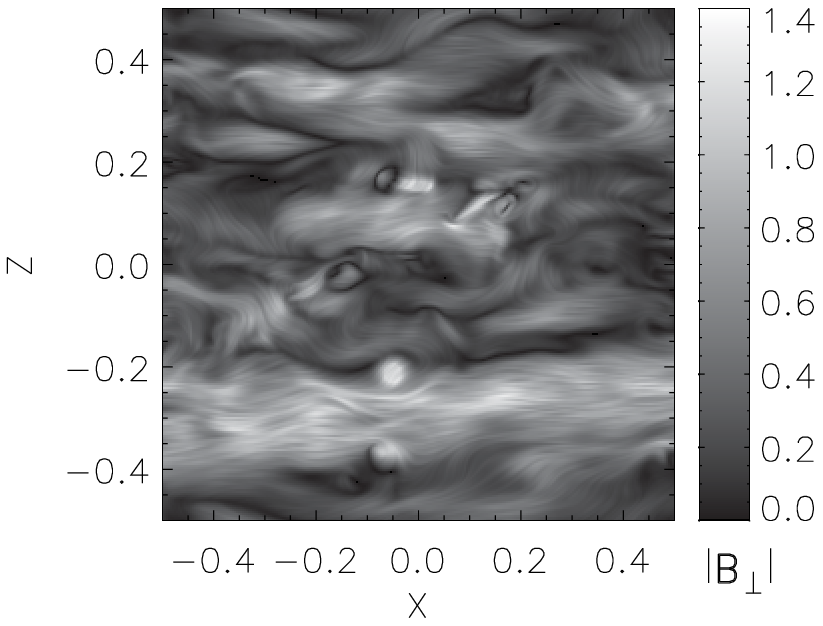}
\includegraphics[width=0.33\textwidth]{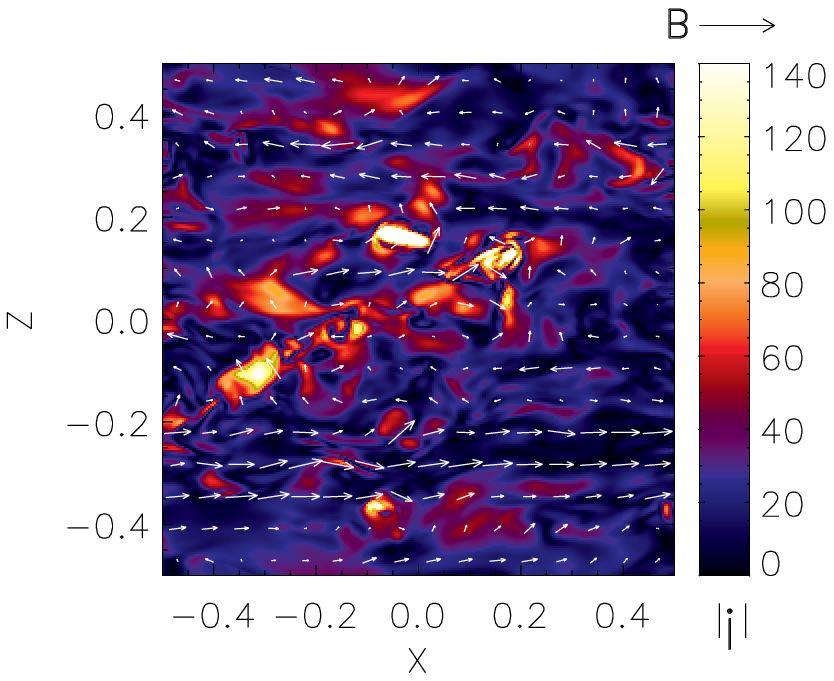}
\caption{Topology and strength of the velocity field (left panel) and magnetic
field (middle panel) in the presence of fully developed turbulence for an
example model with new driving at time $t=10$.  In the right panel we show
distribution of the absolute value of current density $|\vec{J}|$ overlapped
with the magnetic vectors.  The images show the XY-cut (upper row) and XZ-cut
(lower row) of the domain at the midplane of the computational box.  Turbulence
is injected with power $P_{inj}=1$ at scale $k_{inj}=8$ directly in magnetic
field. \label{fig:top_new}}
\end{figure*}

In Figure~\ref{fig:top_new} we show similar examples of XY-cuts (upper row) and
XZ-cuts (lower row) through the box but for a model with the new way of driving
turbulence.  Here, a big number of individual eddies is injected in the magnetic
field with random locations and random orientations in domain. Comparing to
plots in Figure~\ref{fig:top_turb} we see differences but also some clear
similarities.  Among the similarities, we note a highly turbulent region near
the current sheet seen in all XY-cuts, with the current sheet itself strongly
deformed and fragmented into many small scale current sheets (the right column
of Figures~\ref{fig:top_turb} and \ref{fig:top_new}).  We see also some small
increase of magnetic field strength near the current sheet (middle top panels)
resulting from working turbulence in the injection region.  Among the
differences we can list somewhat different distributions of the fragmented
current sheets in the new model with clear enhancements in the locations where
the magnetic eddies are injected at that moment.  These enhancements are clearly
seen in the magnetic topology and current density plots (middle and right
columns).  In order to decrease those strong disturbances of the magnetic lines,
we shall reproduce the same model with higher injection rate and reduced
amplitudes of individual eddies.  Another difference is the strength of current
density $\vc{J}$.  In the model with old driving we see more volume in which
$|\vc{J}|$ reaches high magnitude and its structure is elongated with the local
field.  In the model with new driving, the current density with high strength
seems to be less correlated with the local field, probably due to the presence
of newly injected eddies.  In the intermediate strengths, the structure of
$|\vc{J}|$ seems to be better correlated with the local field.

We see from this comparison that models with different driving of turbulence
demonstrate different topologies of the fields.  In the next sections we show,
that the averaged reconnection rates do not change significantly, confirming
that the way we inject turbulence is of less importance and only its strength
and injection scale have influence on $V_{rec}$.

\subsection{Dependence on Turbulence Strength}
\label{ssec:power}

\begin{figure}
\center
\includegraphics[width=\columnwidth]{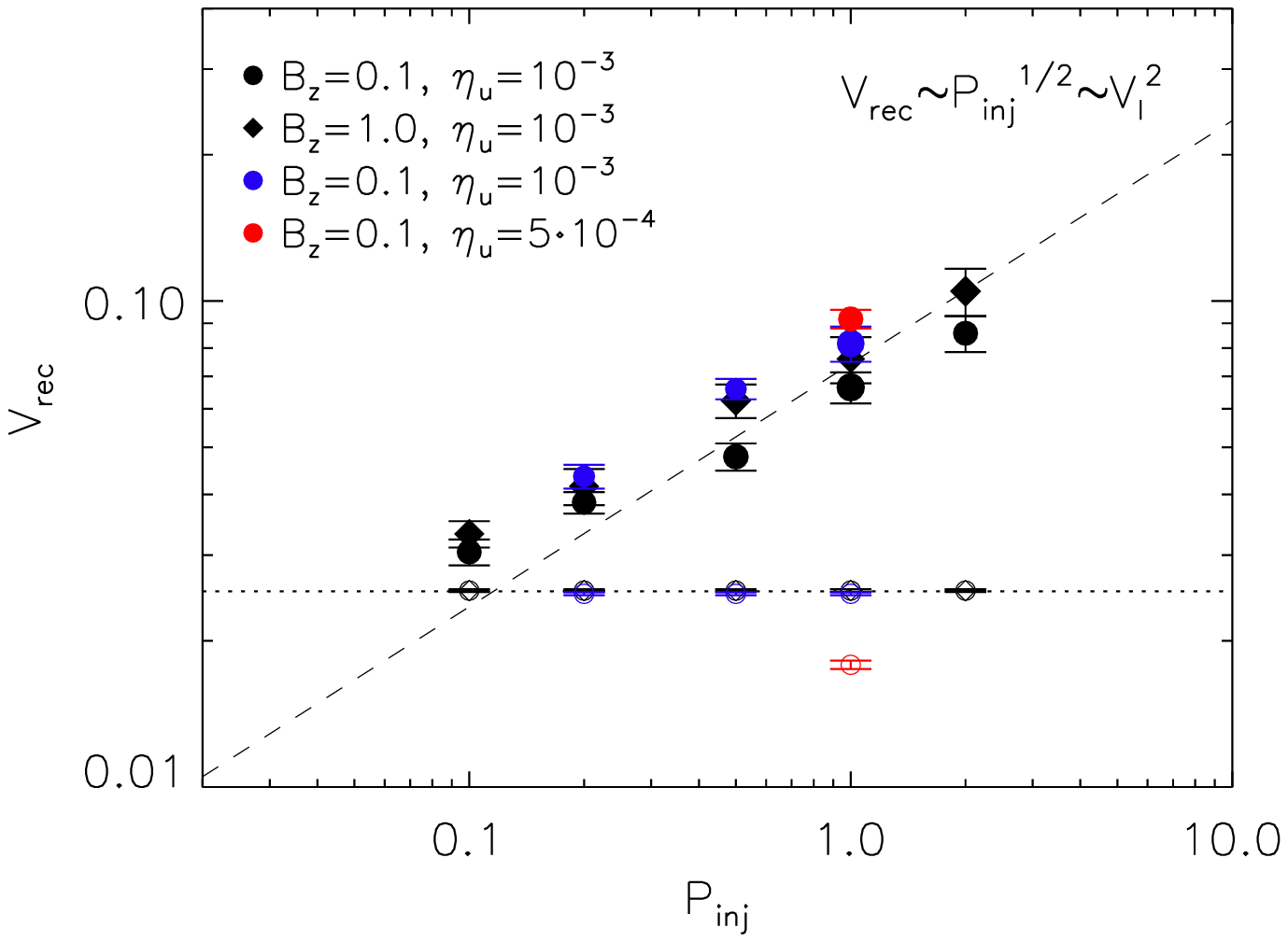}
\caption{The dependence of the reconnection speed $V_{rec}$ on $P_{inj}$ updated
by symbols from new models.  Blue symbols show models with new driving in which
the eddies where injected in magnetic field instead of velocity, as in the
previous models (black symbols).  The dotted line corresponds to the
Sweet-Parker reconnection rate for models with $\eta_{u}=10^{-3}$.  A unique red
symbol shows the reconnection rates from model with new driving in velocity
performed with higher resolution (512x1024x512) and resistivity coefficient
reduced to $\eta_{u}=5\cdot10^{-4}$. Error bars represent the time variance of
$V_{rec}$.  The size of symbols corresponds to the error of $V_{rec}$ (the way
we calculate errors is described in \S\ref{ssec:evolution}).
\label{fig:pow_dep}}
\end{figure}

Models with the new method of turbulence driving are listed in
Table~\ref{tab:models}.  We run a few models with varying turbulent powers in
order to verify if the new driving modifies our previous results.  In these
models we kept the same parameters as in the previous ones, which allowed to
confirm the dependence of the reconnection rate $V_{rec}$ on the power of
injected turbulence $P_{inj}$.

Figure~\ref{fig:pow_dep} shows the values of reconnection speed $V_{rec}$ in
models with turbulent power $P_{inj}$ varying in the range of values by more
than one order of magnitude, from 0.1 to 2.0, for all previously shown models
(black symbols) in \cite{kowal09} and for new models (blue and red symbols) in
which we drove turbulence using the new method.  Because the evolution of
$V_{rec}$ in new models reaches stationarity after time $t=6$, we averaged
$V_{rec}$ from $t=6$ to $t=10$ in these models.  Figure \ref{fig:rate} shows
that the reconnection rates oscillate around their mean values.  In
Figure~\ref{fig:pow_dep} we plot how the averaged reconnection speed depends on
the strength of turbulence.  Filled symbols represent the averaged reconnection
rate in the presence of turbulence.  The dotted line corresponds to the
reconnection rate during the Sweet-Parker process, i.e. without turbulence.  The
error bars show the time variance of $V_{rec}$.  The size of symbols indicates
the uncertainty in our estimate of the reconnection speed $\Delta V_{rec, LV}$
normalized to the uncertainty in the reconnection speed during the Sweet-Parker
evolution $\Delta V_{rec, SP}$.  It is calculated from a formula $size = 2.0 -
\ln \Delta V_{rec, LV} / \ln \Delta V_{rec, SP}$.  If $\Delta V_{rec, LV}$ is of
the order of $\Delta V_{rec, SP}$ their symbols have the same sizes.

The reconnection rates for models with new driving, which is described in
\S\ref{ssec:forcing}, confirm the theoretical dependence of $V_{rec}$ on the
injected power, which scales as $\sim P_{inj}^{1/2}$.  There is no significant
difference between models in which turbulence was driven in velocity and in
magnetic field.  This is in agreement with the LV99 prediction, that the
reconnection rate does not depend on the type of turbulence.

\subsection{Dependence on Injection Scale}
\label{ssec:injection_scale}

The reconnection rate $V_{rec}$ in the presence of turbulence depends only on
the strength of turbulence and its injection scale $l_{inj}$, according to
Equation~\ref{eq:constraint}, for a fixed magnitude of the antiparallel magnetic
field component.  In the previous subsection we presented studies on the
turbulent power dependence.  In this subsection we aim to study the injection
scale dependence.  For this purpose we performed several models with the new way
of driving turbulence, as well, in order to verify if they confirm the
dependence of the reconnection speed $V_{rec}$ on the scale $l_{inj}$ at which
we inject turbulence.  The new models are listed in Table~\ref{tab:models}.  We
inject turbulence at several scales, from $k_{inj} = 8$ to $k_{inj} = 32$.  At
the upper end of this range the turbulence barely broadens the Sweet-Parker
current sheet. At the lower end the turbulent eddies are barely contained within
the volume in which we excite turbulent motions.

\begin{figure}
\center
\includegraphics[width=\columnwidth]{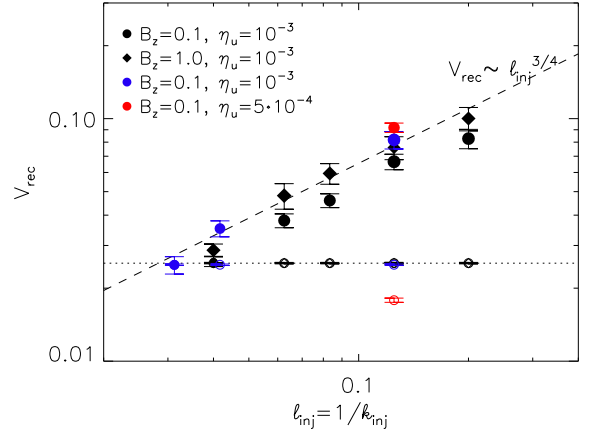}
\caption{The dependence of the reconnection speed $V_{rec}$ on $l_{inj}$ with
additional models in which turbulence was driven in a new way, as described in
\S\ref{ssec:forcing}.  Similarly to Fig.~\ref{fig:pow_dep}, blue symbols show
models with perturbed magnetic field, and red symbols correspond to a high
resolution model with reduced uniform resistivity in which turbulence was driven
in velocity.  The dotted line corresponds to the Sweet-Parker reconnection rate
for models with $\eta_{u}=10^{-3}$.  Error bars and the size of symbols have the
same meaning as in Fig.~\ref{fig:pow_dep}.  \label{fig:sca_dep}}
\end{figure}

In Figure~\ref{fig:sca_dep} we present the reconnection speed dependence on the
injection scale.  We plot the averaged $V_{rec}$ for old models (black symbols)
completed by the values from new models with alternative driving (blue and red
symbols).  From the plot we clearly see a strong dependence of the reconnection
rate on the injection scale.  The new models very precisely follow the same
dependence, confirming again that the type of turbulent driving has no influence
on the process of reconnection, and only the power and injection scale of this
driving have strong importance.  Similarly, as in the power dependence plot, the
new models have slightly higher reconnection speeds comparing to the old ones.
This is due to reduced numerical dissipation of velocity, since in the new
models we used higher order methods.  Dissipation removes energy at small
scales.  If it is smaller, due to higher order numerical scheme, the turbulent
fluctuations reach higher amplitudes at the current sheet scale.  This
influences the rate of individual reconnection events improving slightly the
global reconnection rate $V_{rec}$.

Figure~\ref{fig:sca_dep} shows a bit weaker scaling with the injection scale
than that predicted by LV99 model, i.e. $V_{rec} \sim l_{inj}$.  We see several
possible sources for the discrepancy.  For instance, the existence of a
turbulent inverse cascade can modify the effective $l_{inj}$.  In addition,
reconnection can also modify the characteristics of turbulence, such as the
power spectrum and anisotropy.  We aim to study these problems in the future
work.

\subsection{Dependence on Viscosity}
\label{ssec:viscosity}

In \cite{kowal09} we performed studies of the reconnection rate on the
resistivity, both the uniform and anomalous ones, and we obtained great
agreement with the Sweet-Parker scaling $V_{rec} \sim \eta_u^{1/2}$ for the case
without turbulence, and no dependence on resistivity in the presence of
turbulence, as was predicted in LV99.  In this section we performed additional
studies of the reconnection rate dependence on viscosity.  The dissipation scale
of turbulent cascade is related to the magnitude of viscosity.  If the
dissipation works at scales larger than the current sheet thickness, the
turbulence cascade stops before reaching the current sheet and the global
reconnection rate should be reduced.  The reconnection will be enhanced still by
the broadened ejection region, allowing for more efficient removal of the
reconnected magnetic flux.

\begin{figure}
\center
\includegraphics[width=\columnwidth]{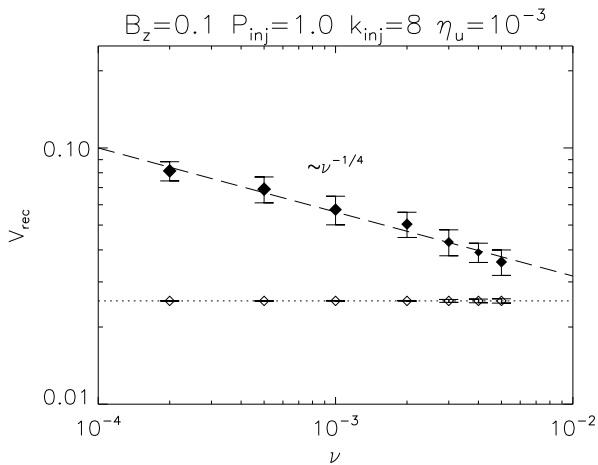}
\caption{The dependence of the reconnection speed $V_{rec}$ on the uniform
viscosity coefficient $\nu$.  As explained in the text, the reconnection speed
is reduced with increasing value of $\nu$.  The dotted line corresponds to the
Sweet-Parker reconnection rate.  Error bars and the size of symbols have the
same meaning as in Fig.~\ref{fig:pow_dep}. \label{fig:vis_dep}}
\end{figure}

In Figure~\ref{fig:vis_dep} we show reconnection rates for models with varying
viscosity coefficient.  Although there is not prediction for this dependence in
the LV99 model, we could test it numerically.  In the Figure~\ref{fig:vis_dep}
we see a weak dependence $V_{rec} \sim \nu^{-1/4}$.  This dependence might be
also useful in understanding the reconnection speed differences between models
with the same set of parameters but different resolutions, or solved with
different orders of the numerical scheme.  At low resolutions or low order
schemes, the numerical viscosity is expected to be larger, thus we should
observe reduced reconnection speeds in those cases.  This is confirmed in
Figures~\ref{fig:pow_dep} and \ref{fig:sca_dep} where we compare old models done
with the second order scheme, and new models done with higher order schemes and
higher resolutions.  In those plots all new models demonstrate slightly higher
reconnection rates.

\section{Discussion}
\label{sec:discussion}

\subsection{LV99 in Collisional and Collisionless Plasma}

The LV99 model was introduced for both collisional and collisionless media and
it claimed that the microphysics of collisionless reconnection events does not
change the resulting reconnection rates.  This point was subjected to further
scrutiny in \cite{eyink11} who provided a thorough investigation of the problem
and concluded that for most of astrophysical collisionless plasmas the LV99
model should be applicable, provided that plasma is turbulent.  With turbulence
being ubiquitous in astrophysical conditions, this hardly constraints the
applicability of the LV99 model.

The LV99 model of reconnection is applicable to the collisional medium, such as
the ISM, which is both turbulent and magnetized, and where the Hall-MHD
reconnection does not work \citep{yamada07}.  For instance, for Hall-MHD
reconnection to be applicable, it is required that the Sweet-Parker current
sheet $\delta_{SP}$ width is smaller than the ion inertial length $d_i$. Thus,
the ``reconnection criterion for media to be collisionless'' is
$(L/d_i)^{1/2}/(\omega_c \tau_e)<1$, which presents a much severe constraint on
the possible rate of collisions.  As a result magnetic reconnection happens to
be mediated by the Hall-MHD only if the extend of the contact region $L$ (see
Fig.~\ref{fig:lv99model}) does not exceed $10^{12}$~cm. Magnetic fields in the
ISM should interact over much larger scales.

The LV99 model works in astrophysical environments to which the Hall-MHD
reconnection is applicable, as well, like Solar corona, interplanetary medium,
if the level of turbulence is high enough.  The reconnection on microscales can
happen fast, i.e. in the Hall-MHD fashion.  This may not change, however, the
global reconnection rate.  The LV99 model shows that even with relatively slow
Sweet-Parker reconnection at microscales the global reconnection is limited not
by Ohmic resistivity, but by the rate of magnetic field wondering.  We believe
that the Hall-MHD {\it local} reconnection of magnetic fields is taking place in
the interplanetary medium, which is being tested by local {\em in-situ}
measurements, while the {\it global} reconnection rates are determined by
magnetic field wandering as prescribed in LV99.

\subsection{Limitations of 2D Reconnection}

In the absence of quantitative model to be tested, simulations aimed at studying
reconnection speed have been done in 2D, both for collisional and collisionless
regimes.  This allowed to achieve higher resolutions (compared to those
contemporary available in 3D) but substantially constrained magnetic field
dynamics.  For instance, the closest study to ours was done by
\cite{matthaeus85} \cite[see also][]{matthaeus86}.  The authors studied 2D
magnetic reconnection in the presence of external turbulence.  An enhancement of
the reconnection rate was reported, but the numerical setup precluded the
calculation of a long term average reconnection rate.  A more recent study along
the approach in \cite{matthaeus85} is one in \cite{watson07}, where the effects
of small scale turbulence on 2D reconnection were studied and no significant
effects of turbulence on reconnection were reported for the setup chosen by the
authors. Later, \cite{servidio10} redone the modeling of 2D turbulent
reconnection following \cite{matthaeus85} with much higher resolutions.  They
used an advanced technique to detect precisely all X-points in the domain and
then performed statistical studies confirming the Sweet-Parker relation for the
reconnection rate as a function of X-point geometry.  The development of
different techniques to study magnetic reconnection is very important.  Even
though their modeling was limited to one type of highly superAlfv\'enic decaying
turbulence (the initial uniform magnetic field was zero), they reported
reconnection rates with normalized values $0.1-0.3$ and confirmed the importance
of turbulence for modifying the character of magnetic reconnection and specifies
heating and transport as the effect of particular significance, as well as
formation of Petschek-type ``X-points'' in 2D turbulence.  Due to the lack of
large scale magnetic field configuration, their model represents a specific
case, far from generic situation observed in the astrophysical objects where the
mean and turbulent components of magnetic fields have comparable strengths.
Therefore, these studies cannot predict the global reconnection rate, as well.
Moreover, \cite{servidio10} interpreted successful numerical confirmation of the
LV99 model as a result of strong turbulence, although \cite{kowal09} addressed
this problem carefully showing that the amplitudes of velocity fluctuations,
both injected and obtained from spectra of developed and stationary turbulence,
are fractions of Alfv\'en speed.

The fact that our study is in 3D is essential, as the LV99 model is
intrinsically three dimensional.  The general picture is of tangled field lines
with reconnection taking place via a series of ``Y-points'' or modified
Sweet-Parker sheets distributed in some fractal way throughout the turbulence. A
large scale Sweet-Parker sheet will be replaced by a more fractured surface, but
the current sheets will occupy a vanishingly small fraction of the total volume
and the field reversal will remain relatively well localized.  The model
predicts that the reconnection speed would be approximately equal to the strong
turbulent velocity with a modest dependence on the ratio of the eddy length to
the current sheet length.  There should be no dependence on resistivity.  The
major results contained in our figures showing the dependence of the
reconnection speed on resistivity, input power and input scale agree with the
quantitative predictions of the LV99 model.  We are not aware of any competing
models to compare our simulations with.

The major differences from the present study stem from the fact that we test a
3D model of reconnection, as the LV99 depends on effects, like field wandering,
that happen only in 3D.  In order to show how different 2D and 3D worlds are, we
performed similar studies to those presented in \cite{kowal09}, but limiting the
domain to two dimensions \cite[see][]{kulpa10}.  In \cite{kulpa10} we
demonstrated that 2D magnetic reconnection in the presence of turbulence depends
on the Ohmic resistivity, therefore, it is not fast.  Also, the dependencies on
the turbulent power and injection scales were significantly weaker than in the
LV99.  This dependence of 2D reconnection rate on Ohmic resistivity in the
presence of turbulence, although weaker than the Sweet-Parker relation $V_{rec}
\sim \eta^{-1/2}$, has been independently confirmed by \cite{loureiro09}
studies, performed with very different approach.  These differences call for
deliberation with simple extension of conclusions coming from the 2D modeling to
natural for magnetic field fully three dimensional world.

\subsection{Applications of the LV99 Model}

Reconnection is one of the most fundamental processes involving magnetic fields
in conducting fluids or plasmas.  Therefore, the identification of a robust
process responsible for reconnection has many astrophysically important
consequences.  Below we list a few selected implications of the successful
validation of the LV99 model.

Numerical studies on Fermi acceleration in turbulent reconnection have a long
history \cite[e.g.][]{matthaeus84,goldstein86,ambrosiano88,drake06,hoshino12}.
In the Sweet-Parker model, it has been shown that particles can accelerate due
to the induced electric field in the reconnection zone \citep{litvinenko03}.
This $one-shot$ acceleration process, however, is constrained by the narrow
thickness of the acceleration zone which has to be larger than the particle
Larmor radius and by the strength of the magnetic field.  Therefore, the
efficiency of this process is rather limited.  Besides, it also does not predict
a power-law spectrum, as generally observed for cosmic rays.  Observations have
always been suggestive that magnetic reconnection can happen at a high speed in
some circumstances, in spite of the theoretical difficulties in explaining it.
For instance, the phenomenon of solar flares suggests that magnetic reconnection
should be first slow in order to ensure the accumulation of magnetic flux and
then suddenly become fast in order to explain the observed fast release of
energy.  The LV99 model can naturally explain this and other observational
manifestations of magnetic reconnection.  Consider a particle entrained on a
reconnected magnetic field line (see Fig.\ref{fig:lv99model}).  This particle
may bounce back and forth between magnetic mirrors formed by oppositely directed
magnetic fluxes moving towards each other with the velocity $V_{rec}$.  Each
bounce will increase the energy of a particle in a way consistent with the
requirements of the first-order Fermi process\footnote{Another way of
understanding the acceleration of energetic particles in the reconnection
process above is to take into account that the length of magnetic field lines is
decreasing during reconnection.  As a result, the physical volume of the
energetic particles entrained on the field lines is shrinking.  Thus, due to
Liouville's theorem, their momentum should increase to preserve the constancy of
the phase volume.} \citep{dalpino01,dalpino03,dalpino05,lazarian06}.  This is in
contrast to the second-order Fermi acceleration that is frequently discussed in
terms of accelerating particles by turbulence generated by reconnection
\citep{larosa06}.  The numerical studies of the particle acceleration supporting
these ideas have been already started \citep{kowal11a,kowal11b}.  An interesting
property of this acceleration mechanism is that it is also potentially testable
observationally, since the resulting spectrum of accelerated particles is
different from that arising from a shock.  \cite{dalpino01,dalpino05} used this
mechanism of particle acceleration to explain the synchrotron power-law spectrum
arising from the flares of the microquasar GRS 1915+105.

Further applications can be found in the solar physics.  Following
\cite{zweibel09} we note that solar flares inspired much of the earlier research
on reconnection \cite[see][]{pneuman81,bastian98}.  As the plasma involved is
substantially rarefied, the restrictive conditions for the collisionless
reconnection are satisfied in this particular environment.  \cite{cassak05}
stated that bistable Hall reconnection can be important in this case. Stochastic
reconnection provides an alternative explanation.  Indeed, an important
prediction of the LV99 model is related to the {\it reconnection instability}
that arises in the situation when the initial structure of the flux prior to
reconnection is laminar.  Reconnection at the Sweet-Parker rate is negligible.
This allows magnetic flux to accumulate.  However, when the degree of
stochasticity exceeds a threshold value, the reconnection itself should excite
more turbulence, creating a positive feedback resulting in a flare
\cite[see][]{lazarian09b}.  The instability is a generic property of laminar
field reconnection in both collisionless and collisional environments.
Referring to the Sun, one may speculate that the difference between gradual and
eruptive flares arises from the original state of magnetic field prior to the
flare, at least in some specific situations.  In the case when the magnetic
field is sufficiently turbulent the accumulation of magnetic flux does not
happen and the flare is gradual.  Similarly, the observed spatial spread of
energy release during solar flares may be due to the spread of the region of
turbulent fields once reconnection is initiated at one place.  Recent
observations demonstrate that gradual flares occur rather in regions with large
scale and weak magnetic fields for which Alfv\'en times are large
\citep{shibata11}.  In light of that, the difference in Alfv\'en times may
explain different time scales in gradual and impulsive flares.  Further research
is necessary for establishing the role of turbulence in changing the time scale
of flare evolution.

The LV99 model can find its application in the removal of magnetic flux from
the star formation regions.  \cite{shu06} showed that magnetic field is removed
from the star forming core cluster faster than it is allowed by the standard
ambipolar diffusion scenario \citep{tassis05a,tassis05b}.  \cite{shu07} proposed
a mechanism using efficient reconnection through ``hyper-resistivity''.
\cite{santos10} performed numerical studies of such a concept, replacing the
``hyper-resistivity'' with efficient stochastic reconnection.  They reported
removal of strong anticorrelations  of magnetic field through ``reconnection
diffusion'', which can mimic the effect of enhanced Ohmic resistivity.

LV99 showed that fast reconnection of stochastic magnetic field makes the models
of strong MHD turbulence self-consistent, because the critical balance in the
GS95 model requires the existence of eddy-type motions perpendicular to the
magnetic field.  In the absence of reconnection this would result in unresolved
knots that should drain energy from the cascade.  The estimates in LV99 showed
that the rates of reconnection predicted by the model are sufficient to resolve
magnetic knots within one period.

\subsection{Reasons for Slow Adaptation of the LV99 Model}

The LV99 model of magnetic reconnection in the presence of weakly stochastic
magnetic fields was proposed by \citeauthor{lazarian99} in
\citeyear{lazarian99}.  However, due to a few objective factors it met less
enthusiasm in the community than, for example, the X-point collisionless
reconnection.
We believe that there were three major factors responsible for this.
\begin{enumerate}
 \item The collisionless X-point reconnection was initiated and supported by
numerical simulations, while LV99 was a theory.  Its numerical testing became
possible only recently.  The reconnection subject had a history of failed
theories and models, which without direct numerical support were taken with a
grain of salt.
 \item The acceptance of the idea of astrophysical fluids generically being in
turbulent state had much less observational support at that time compared to the
present day.  By now we have much more evidence which allows to claim that
models not taking the pre-existent turbulence has little relevance to
astrophysics.
 \item The analytical solutions of LV99 were based on the use of GS95 model of
turbulence.  The GS95 model of turbulence, in fact, was extended LV99 by
introducing the concept of local reference frame for turbulent eddies and by
extending the GS95 scalings to the subAlfv\'enic case.  The GS95 theory was far
from being generally accepted at the time of the LV99 publishing.
\end{enumerate}

The situation has changed substantially by now.  First of all, GS95 was
successfully tested numerically \citep{cho00,maron01,cho02b} and their ideas
have been extended to describing the Alfv\'enic cascade in compressible MHD
turbulence \cite[see][]{cho02a,cho03,kowal10}\footnote{There are attempts to
modify GS95 theory by supplementing it with additional effects, like dynamical
alignment \citep{boldyrev05,boldyrev06}, polarization \citep{beresnyak06},
non-locality \cite{gogoberidze07}.  All these attempts, however, do not change
the very nature of the GS95 model.  Moreover, some recent studies
\cite{beresnyak09,beresnyak10,beresnyak11} indicate that the numerical
motivation for introducing these attempts may be due to the insufficient
inertial range of the simulations involved.}.  The so-called ``Big Power Law in
the Sky'' indicating the presence of turbulence on scales from tens of parsecs to
thousands of kilometers has been extended \citep{chepurnov10a}, and the
observations of gas and synchrotron emission provided extended number of direct
turbulence measurements confirming their presence
\cite[see][]{burkhart10,gaensler11}.  Finally, the situation has changed with
the numerical testing of the LV99 model.  The 3D MHD simulations in
\cite{kowal09} supported the predictions in the LV99 paper and our present work
goes further with testing this model, by including different types of energy
injection.

It is worth noting also, that there is some implicit observational evidence in
the favor of the LV99 model, like observations of the thick reconnection current
outflow regions observed in the Solar flares \citep{ciaravella08}.
\cite{sych09}, explaining quasi-periodic pulsations in observed flaring energy
releases at an active region above the sunspot, proposed that the wave packets
arising from the sunspots can trigger such pulsations.  They established a
phenomenological relation between oscillations in a sunspot and pulsations in
flaring energy releases.  This phenomenon can be naturally explained by the LV99
model.

\conclusions
\label{sec:summary}

In this article we performed additional testing of the LV99 model of fast
reconnection under different types of turbulent driving using 3D numerical
simulations.  We have introduced a new method of driving turbulence by direct
injection of the velocity or magnetic eddies with random locations in the
domain.  We analyzed the dependence of the reconnection speed on the turbulence
injection power, on the injection scale, as well as on the viscosity.  We found
that:
\begin{itemize}
\item We observe similar changes of the topology of the magnetic field near the
interface of oppositely directed magnetic field lines in models with two
different turbulence injection mechanisms.  These changes include the
fragmentation of the current sheet, favoring multiple simultaneous reconnection
events, as well as a substantial increase in the thickness of the outflow of
reconnected magnetic flux and matter.

\item The relation between the reconnection rate $V_{rec}$ and turbulent power
$P_{inj}$ remains unchanged under two different mechanisms of energy injection
and is confirmed by new models to be $V_{rec} \sim P_{inj}^{1/2}
\sim V_l^2$, in agreement with the LV99 prediction.  Moreover, the injection in
magnetic field produces similar effects on the reconnection as injection in
velocity, remaining the dependence unaltered.

\item The reconnection rate grows with the size of the injected eddies, which
can be directly related to the turbulence injection scale.  The rate of growth,
for the models with old and new driving mechanism, is approximated by $V_{rec}
\sim l_{inj}^{3/4}$ scaling which agrees with the previously obtained scaling.
Somewhat steeper LV99 prediction, $V_{rec}\sim l_{inj}$, could results from
limitations in the dynamic range available for study.

\item Reconnection in the presence of weak turbulence is only weakly sensitive
to viscosity $\nu$.  From performed simulations we obtained a dependence
$V_{rec} \sim \nu^{-1/4}$ for one set of parameters: $P_{inj}=1.0$, $k_{inj}=8$,
$\eta_u=10^{-3}$, $B_{0z}=0.1$.
\end{itemize}

\begin{acknowledgements}
The research of GK is supported by FAPESP grant no. 2009/50053-8, AL is
supported by the Center for Magnetic Self-Organization in Laboratory and
Astrophysical Plasmas, NSF grant AST-08-08118 and NASA grant NNX09AH78G.  The
work of ETV is supported by the National Science and Engineering Research
Council of Canada.  The research of KOM is supported by the Polish Ministry of
Science and Higher Education through grants: 92/N-ASTROSIM/2008/0 and
3033/B/H03/2008/35.  This research also was supported by the National Science
Foundation project TG-AST080005N through TeraGrid resources provided by Texas
Advanced Computing Center (TACC:http://www.tacc.utexas.edu).  Part of this work
was made possible by the facilities of the Shared Hierarchical Academic Research
Computing Network (SHARCNET:http://www.sharcnet.ca) and the GALERA supercomputer
in the Academic Computer Centre in Gda\'{n}sk (TASK:http://www.task.gda.pl).
\end{acknowledgements}



\begin{thebibliography}{}

 \bibitem[Alvelius(1999)]{alvelius99}
  Alvelius, K.: Random forcing of three-dimensional homogeneous turbulence,
  \pof, 11, 1880--1889, 1999.

 \bibitem[Ambrosiano et al.(1988)]{ambrosiano88}
  Ambrosiano, J., Matthaeus, W.~H., Goldstein, M.~L., and Plante, D.: Test
  particle acceleration in turbulent reconnecting magnetic fields, \jgr, 93,
  14383--14400, 1988.

 \bibitem[Armstrong et al.(1995)]{armstrong95}
  Armstrong, J.~W., Rickett, B.~J., and Spangler, S.~R.: Electron density
  power spectrum in the local interstellar medium, \apj, 443, 209--221, 1995.

 \bibitem[Balbus and Hawley(1998)]{balbus98}
  Balbus, S.~A.~and Hawley, J.~F.: Instability, turbulence, and enhanced
  transport in accretion disks, \rmp, 70, 1--53, 1998.

 \bibitem[Bastian et al.(1998)]{bastian98}
  Bastian, T.~S., Benz, A.~O., and Gary, D.~E.: Radio Emission from Solar
  Flares, \araa, 36, 131--188, 1998.

 \bibitem[Beck(2002)]{beck02}
  Beck, R.: Magnetic Fields in Spiral Arms and Bars, Disks of Galaxies:
  Kinematics, Dynamics and Peturbations, 275, 331--342, 2002.

 \bibitem[Beresnyak(2011)]{beresnyak11}
  Beresnyak, A.: Spectral Slope and Kolmogorov Constant of MHD Turbulence,
  \prl, 106, 075001, 2011.

 \bibitem[Beresnyak and Lazarian(2006)]{beresnyak06}
  Beresnyak, A.~and Lazarian, A.: Polarization Intermittency and Its
  Influence on MHD Turbulence, \apjl, 640, L175--L178, 2006.

 \bibitem[Beresnyak and Lazarian(2009)]{beresnyak09}
  Beresnyak, A.~and Lazarian, A.: Comparison of Spectral Slopes of
  Magnetohydrodynamic and Hydrodynamic Turbulence and Measurements of
  Alignment Effects, \apj, 702, 1190--1198, 2009.

 \bibitem[Beresnyak and Lazarian(2010)]{beresnyak10}
  Beresnyak, A.~and Lazarian, A.: Scaling Laws and Diffuse Locality of
  Balanced and Imbalanced Magnetohydrodynamic Turbulence, \apjl, 722,
  L110--L113, 2010.

 \bibitem[Bhattacharjee et al.(2003)]{bhattacharjee03}
  Bhattacharjee, A., Ma, Z.~W., and Wang, X.: Recent Developments in
  Collisionless Reconnection Theory: Applications to Laboratory and
  Astrophysical Plasmas, \lnp, 614, 351--375, 2003.\

 \bibitem[Bhattacharjee et al.(2009)]{bhattacharjee09}
  Bhattacharjee, A., Huang, Y.-M., Yang, H., and Rogers, B.: Fast reconnection
  in high-Lundquist-number plasmas due to the plasmoid Instability, \pop, 16,
  112102, 2009.

 \bibitem[Biskamp(1996)]{biskamp96}
  Biskamp, D.: Magnetic Reconnection in Plasmas, \apss, 242, 165--207, 1996.

 \bibitem[Biskamp(2000)]{biskamp00}
  Biskamp, D.: Magnetic Reconnection in Plasmas, Cambridge University Press,
  Cambridge, UK, 2000.

 \bibitem[Boldyrev(2005)]{boldyrev05}
  Boldyrev, S.: On the Spectrum of Magnetohydrodynamic Turbulence, \apjl,
  626, L37--L40, 2005.

 \bibitem[Boldyrev(2006)]{boldyrev06}
  Boldyrev, S.: Spectrum of Magnetohydrodynamic Turbulence, \prl, 96, 115002,
  2006.

 \bibitem[Burkhart et al.(2010)]{burkhart10}
  Burkhart, B., Stanimirovi{\'c}, S., Lazarian, A., and Kowal, G.:
  Characterizing Magnetohydrodynamic Turbulence in the Small Magellanic
  Cloud, \apj, 708, 1204--1220, 2010.

 \bibitem[Cassak et al.(2005)]{cassak05}
  Cassak, P.~A., Shay, M.~A., and Drake, J.~F.: Catastrophe Model for Fast
  Magnetic Reconnection Onset, \prl, 95, 235002, 2005.

 \bibitem[Chepurnov and Lazarian(2010)]{chepurnov10a}
  Chepurnov, A.~and Lazarian, A.: Extending the Big Power Law in the Sky with
  Turbulence Spectra from Wisconsin H{$\alpha$} Mapper Data, \apj, 710,
  853--858, 2010.

 \bibitem[Chepurnov et al.(2010)]{chepurnov10b}
  Chepurnov, A., Lazarian, A., Stanimirovi{\'c}, S., Heiles, C., and Peek,
  J.~E.~G.: Velocity Spectrum for H I at High Latitudes, \apj, 714,
  1398--1406, 2010.

 \bibitem[Cho and Lazarian(2002)]{cho02a}
  Cho, J.~and Lazarian, A.: Compressible Sub-Alfv{\'e}nic MHD Turbulence in
  Low-{$\beta$} Plasmas, \prl, 88, 245001, 2002.

 \bibitem[Cho and Lazarian(2003)]{cho03}
  Cho, J.~and Lazarian, A.: Compressible magnetohydrodynamic turbulence: mode
  coupling, scaling relations, anisotropy, viscosity-damped regime and
  astrophysical implications, \mnras, 345, 325--339, 2003.

 \bibitem[Cho et al.(2002)]{cho02b}
  Cho, J., Lazarian, A., and Vishniac, E.~T.: Simulations of
  Magnetohydrodynamic Turbulence in a Strongly Magnetized Medium, \apj, 564,
  291--301, 2002.

 \bibitem[Cho and Vishniac(2000)]{cho00}
  Cho, J.~and Vishniac, E.~T.: The Anisotropy of Magnetohydrodynamic
  Alfv{\'e}nic Turbulence, \apj, 539, 273--282, 2000.

 \bibitem[Ciaravella and Raymond(2008)]{ciaravella08}
  Ciaravella, A.~and Raymond, J.~C.: The Current Sheet Associated with the
  2003 November 4 Coronal Mass Ejection: Density, Temperature, Thickness, and
  Line Width, \apj, 686, 1372--1382, 2008.

 \bibitem[Crutcher(1999)]{crutcher99}
  Crutcher, R.~M.: Magnetic Fields in Molecular Clouds: Observations Confront
  Theory, \apj, 520, 706--713, 1999.

 \bibitem[de Gouveia Dal Pino and Lazarian(2001)]{dalpino01}
  de Gouveia Dal Pino, E.~M.~and Lazarian, A.: Constraints on the
  Acceleration of Ultra-High-Energy Cosmic Rays in Accretion-induced Collapse
  Pulsars, \apj, 560, 358--364, 2001.

 \bibitem[de Gouveia Dal Pino and Lazarian(2003)]{dalpino03}
  de Gouveia Dal Pino, E.~M.~and Lazarian, A.: The role of Violent Magnetic
  Reconnection on the Production of the Large Scale Superluminal Ejections of
  the Microquasar GRS 1915+105, arXiv:astro-ph/0307054, 2003.

 \bibitem[de Gouveia dal Pino and Lazarian(2005)]{dalpino05}
  de Gouveia dal Pino, E.~M.~and Lazarian, A.: Production of the large scale
  superluminal ejections of the microquasar GRS 1915+105 by violent magnetic
  reconnection, \aap, 441, 845--853, 2005.

 \bibitem[Drake et al.(2006)]{drake06}
  Drake, J.~F., Swisdak, M., Che, H., and Shay, M.~A.: Electron acceleration
  from contracting magnetic islands during reconnection, \nat, 443, 553--556,
  2006.

 \bibitem[Elmegreen and Scalo(2004)]{elmegreen04}
  Elmegreen, B.~G.~and Scalo, J.: Interstellar Turbulence I: Observations and
  Processes, \araa, 42, 211--273, 2004.

 \bibitem[Eyink et al.(2011)]{eyink11}
  Eyink, G.~L., Lazarian, A., and Vishniac, E.~T.: Fast Magnetic Reconnection
  and Spontaneous Stochasticity, \apj, 743, 51, 2011.

 \bibitem[Fitzpatrick(2004)]{fitzpatrick04}
  Fitzpatrick, R.: Scaling of forced magnetic reconnection in the
  Hall-magnetohydrodynamic Taylor problem, \pop, 11, 937--946, 2004.

 \bibitem[Forbes(2000)]{forbes01}
  Forbes, T.~G.: The nature of Petschek-type reconnection, Earth Planets Space, 53, 423, 2001

 \bibitem[Gaensler et al.(2011)]{gaensler11}
  Gaensler, B.~M., Haverkorn, M., Burkhart, B., Newton-McGee, K.~J., Ekers,
  R.~D., Lazarian, A., McClure-Griffiths, N.~M., Robishaw, T., Dickey, J.~M.,
  and Green, A.~J.: Low-Mach-number turbulence in interstellar gas revealed
  by radio polarization gradients, \nat, 478, 214--217, 2011.

 \bibitem[Gogoberidze(2007)]{gogoberidze07}
  Gogoberidze, G.: On the nature of incompressible magnetohydrodynamic
  turbulence, \pop, 14, 022304, 2007.

 \bibitem[Goldreich and Sridhar(1995)]{goldreich95}
  Goldreich, P.~and Sridhar, S.: Toward a theory of interstellar turbulence. 2:
  Strong alfvenic turbulence, \apj, 438, 763--775, 1995. (GS95)

 \bibitem[Goldstein et al.(1986)]{goldstein86}
  Goldstein, M.~L., Matthaeus, W.~H., and Ambrosiano, J.~J.: Acceleration of
  charged particles in magnetic reconnection Solar flares, the magnetosphere,
  and solar wind, \grl, 13, 205--208, 1986.

 \bibitem[Gottlieb et al.(2009)Gottlieb, Ketcheson and Shu]{gottlieb09}
  Gottlieb, S., Ketcheson, D.~I. and Shu, C.-W.: High Order Strong Stability
  Preserving Time Discretizations, \jsc, 38, 251--289, 2009.

 \bibitem[Hanasz et al.(2009)]{hanasz09}
  Hanasz, M., Otmianowska-Mazur, K., Kowal, G., and Lesch, H.:
  Cosmic-ray-driven dynamo in galactic disks. A parameter study, \aap, 498,
  335--346, 2009.

 \bibitem[He et al.(2011)]{he11}
  He, Z., Li, X., Fu, D., and Ma, Y.: A 5th order monotonicity-preserving
  upwind compact difference scheme, Science in China G: Physics and Astronomy,
  54, 511--522, 2011.

 \bibitem[Hoshino(2012)]{hoshino12}
  Hoshino, M.: Stochastic Particle Acceleration in Multiple Magnetic Islands
  during Reconnection, arXiv:1201.0837, 2012.

 \bibitem[Jacobson and Moses(1984)]{jacobson84}
  Jacobson, A.~R.~and Moses, R.~W.: Nonlocal dc electrical conductivity of a
  Lorentz plasma in a stochastic magnetic field, \pra, 29, 3335--3342, 1984.

 \bibitem[Khanna(1998)]{khanna98}
  Khanna, R.: Generation of magnetic fields by a gravitomagnetic plasma
  battery, \mnras, 295, L6--L10, 1998.

 \bibitem[Kim and Diamond(2001)]{kim01}
  Kim, E.-j.~and Diamond, P.~H.: On Turbulent Reconnection, \apj, 556,
  1052--1065, 2001.

 \bibitem[Kotera and Olinto(2011)]{kotera11}
  Kotera, K.~and Olinto, A.~V.: The Astrophysics of Ultrahigh-Energy Cosmic
  Rays, \araa, 49, 119--153, 2011.

 \bibitem[Kowal et al.(2011a)]{kowal11a}
  Kowal, G., de Gouveia Dal Pino, E.~M., and Lazarian, A.:
  Magnetohydrodynamic Simulations of Reconnection and Particle Acceleration:
  Three-dimensional Effects, \apj, 735, 102, 2011.

 \bibitem[Kowal et al.(2011b)]{kowal11b}
  Kowal, G., de Gouveia Dal Pino, E.~M., and Lazarian, A.:
  Acceleration in Turbulence and Weakly Stochastic Reconnection, \prl, 2011, in press.

 \bibitem[Kowal and Lazarian(2010)]{kowal10}
  Kowal, G.~and Lazarian, A.: Velocity Field of Compressible Magnetohydrodynamic
  Turbulence: Wavelet Decomposition and Mode Scalings, \apj, 720, 742--756,
  2010.

 \bibitem[Kowal et al.(2009)]{kowal09}
  Kowal, G., Lazarian, A., Vishniac, E.~T., and Otmianowska-Mazur, K.:
  Numerical Tests of Fast Reconnection in Weakly Stochastic Magnetic Fields,
  \apj, 700, 63--85, 2009.

 \bibitem[Kulpa-Dybe{\l} et al.(2010)]{kulpa10}
  Kulpa-Dybe{\l}, K., Kowal, G., Otmianowska-Mazur, K., Lazarian, A., and
  Vishniac, E.: Reconnection in weakly stochastic B-fields in 2D, \aap, 514,
  A26, 2010.

 \bibitem[Kulpa-Dybe{\l} et al.(2011)]{kulpa11}
  Kulpa-Dybe{\l}, K., Otmianowska-Mazur, K., Kulesza-{\.Z}ydzik, B., Hanasz,
  M., Kowal, G., W{\'o}lta{\'n}ski, D., and Kowalik, K.: Global Simulations
  of the Magnetic Field Evolution in Barred Galaxies Under the Influence of
  the Cosmic-ray-driven Dynamo, \apjl, 733, L18, 2011.

 \bibitem[La Rosa et al.(2006)]{larosa06}
  La Rosa, T.~N., Shore, S.~N., Joseph, T., Lazio, W., and Kassim, N.~E.: The
  Strength and Structure of the Galactic Center Magnetic Field, \jpcs, 54,
  10--15, 2006.

 \bibitem[Lazarian(2006)]{lazarian06}
  Lazarian, A.: Theoretical approaches to particle propagation and
  acceleration in turbulent intergalactic medium, \an, 327, 609, 2006.

 \bibitem[Lazarian(2009)]{lazarian09}
  Lazarian, A.: Obtaining Spectra of Turbulent Velocity from Observations,
  \ssr, 143, 357--385, 2009.

 \bibitem[Lazarian and Pogosyan(1999)]{lazarian99b}
  Lazarian, A.~and Pogosyan, D.: Velocity Modification of HI Spectrum and
  Clouds in Velocity Space, Bulletin of the American Astronomical Society,
  31, 1449, 1999.

 \bibitem[Lazarian and Vishniac(1999)]{lazarian99}
  Lazarian, A.~and Vishniac, E.~T.: Reconnection in a Weakly Stochastic Field,
  \apj, 517, 700--718, 1999. (LV99)

 \bibitem[Lazarian and Vishniac(2009)]{lazarian09b}
  Lazarian, A.~and Vishniac, E.~T.: Model of Reconnection of Weakly
  Stochastic Magnetic Field and its Implications, \rmxac, 36, 81--88, 2009.

 \bibitem[Lazarian et al.(2004)]{lazarian04}
  Lazarian, A., Vishniac, E.~T., and Cho, J.: Magnetic Field Structure and
  Stochastic Reconnection in a Partially Ionized Gas, \apj, 603, 180--197,
  2004.

 \bibitem[Litvinenko(2003)]{litvinenko03}
  Litvinenko, Y.~E.: Particle Acceleration by a Time-Varying Electric Field
  in Merging Magnetic Fields, \solphys, 216, 189--203, 2003.

 \bibitem[Londrillo and Del Zanna(2000)]{londrillo00}
  Londrillo, P.~and Del Zanna, L.: High-Order Upwind Schemes for
  Multidimensional Magnetohydrodynamics, \apj, 530, 508--524, 2000.

 \bibitem[Loureiro et al.(2009)]{loureiro09}
  Loureiro, N.~F., Uzdensky, D.~A., Schekochihin, A.~A., Cowley, S.~C., and
  Yousef, T.~A.: Turbulent magnetic reconnection in two dimensions, \mnras,
  399, L146--L150, 2009.

 \bibitem[Matthaeus et al.(1984)]{matthaeus84}
  Matthaeus, W.~H., Ambrosiano, J.~J., and Goldstein, M.~L.:
  Particle-acceleration by turbulent magnetohydrodynamic reconnection,
  \prl, 53, 1449--1452, 1984.

 \bibitem[Matthaeus and Lamkin(1985)]{matthaeus85}
  Matthaeus, W.~H.~and Lamkin, S.~L.: Rapid magnetic reconnection caused by
  finite amplitude fluctuations, \pof, 28, 303--307, 1985.

 \bibitem[Matthaeus and Lamkin(1986)]{matthaeus86}
  Matthaeus, W.~H.~and Lamkin, S.~L.: Turbulent magnetic reconnection,
  \pof, 29, 2513--2534, 1986.

 \bibitem[Melrose(2009)]{melrose09}
  Melrose, D.~B.: Acceleration Mechanisms, arXiv:0902.1803, 2009.

 \bibitem[Mignone(2007)]{mignone07}
  Mignone, A.: A simple and accurate Riemann solver for isothermal MHD, \jcp,
  225, 1427--1441, 2007.

 \bibitem[Moffat(1978)]{moffat78}
  Moffat, H.~K.: Magnetic Field Generation in Electrically conducting Fluids, 1978, (London, UK/New York, NY:Cambridge University Press).

 \bibitem[Maron and Goldreich(2001)]{maron01}
  Maron, J.~and Goldreich, P.: Simulations of Incompressible Magnetohydrodynamic
  Turbulence, \apj, 554, 1175--1196, 2001.

 \bibitem[Padoan et al.(2006)]{padoan06}
  Padoan, P., Juvela, M., Kritsuk, A., and Norman, M.~L.: The Power Spectrum
  of Supersonic Turbulence in Perseus, \apjl, 653, L125--L128, 2006.

 \bibitem[Padoan et al.(2009)]{padoan09}
  Padoan, P., Juvela, M., Kritsuk, A., and Norman, M.~L.: The Power Spectrum
  of Turbulence in NGC 1333: Outflows or Large-Scale Driving?, \apjl, 707,
  L153--L157, 2009.

 \bibitem[Parker(1957)]{parker57}
  Parker, E.~N.: Sweet's Mechanism for Merging Magnetic Fields in Conducting
  Fluids, \jgr, 62, 509--520, 1957.

 \bibitem[Parker(1992)]{parker92}
  Parker, E.~N.: Fast dynamos, cosmic rays, and the Galactic magnetic field,
  \apj, 401, 137--145, 1992.

 \bibitem[Petschek(1964)]{petschek64}
  Petschek, H.~E.: Magnetic Field Annihilation, NASA Special Publication, 50,
  425--439, 1964.

 \bibitem[Pneuman(1981)]{pneuman81}
  Pneuman, G.~W.: Two-ribbon flares - /Post/-flare loops, Solar Flare
  Magnetohydrodynamics, 379--428, 1981.

 \bibitem[Priest and Forbes(2000)]{priest00}
  Priest, E.~and Forbes, T.: Magnetic Reconnection, Cambridge University Press,
  Cambridge, UK, 2000.

 \bibitem[Santos-Lima et al.(2010)]{santos10}
  Santos-Lima, R., Lazarian, A., de Gouveia Dal Pino, E.~M., and Cho, J.:
  Diffusion of Magnetic Field and Removal of Magnetic Flux from Clouds Via
  Turbulent Reconnection, \apj, 714, 442--461, 2010.

 \bibitem[Schlickeiser and Lerche(1985)]{schlickeiser85}
  Schlickeiser, R.~and Lerche, I.: Cosmic gas dynamics. I - Basic equations
  and the dynamics of hot interstellar matter, \aap, 151, 151--156, 1985.

 \bibitem[Scholer(1989)]{scholer89}
  Scholer, M.: Undriven magnetic reconnection in an isolated current sheet,
  \jgr, 94, 8805--8812, 1989.

 \bibitem[Servidio et al.(2010)]{servidio10}
  Servidio, S., Matthaeus, W.~H., Shay, M.~A., Dmitruk, P., Cassak, P.~A.,
  and Wan, M.: Statistics of magnetic reconnection in two-dimensional
  magnetohydrodynamic turbulence, \pop, 17, 032315, 2010.

 \bibitem[Shay et al.(1998)]{shay98}
 Shay, M.~A., Drake, J.~F., Denton, R.~E., and Biskamp, D.: Structure of the
 dissipation region during collisionless magnetic reconnection, \jgr, 103,
 9165--9176, 1998.

 \bibitem[Shay et al.(2004)]{shay04}
 Shay, M.~A., Drake, J.~F., Swisdak, M., and Rogers, B.~N.: The scaling of
 embedded collisionless reconnection, \pop, 11, 2199--2213, 2004.

 \bibitem[Shibata and Magara(2011)]{shibata11}
  Shibata, K. and Magara, T.: Solar Flares: Magnetohydrodynamic Processes,
  \lrsp, 8, 6--2011.

 \bibitem[Shibata and Tanuma(2001)]{shibata01}
  Shibata, K.~and Tanuma, S.: Plasmoid-induced-reconnection and fractal
  reconnection, \eps, 53, 473--482, 2001.

 \bibitem[Shu et al.(2006)]{shu06}
  Shu, F.~H., Galli, D., Lizano, S., and Cai, M.: Gravitational Collapse of
  Magnetized Clouds. II. The Role of Ohmic Dissipation, \apj, 647, 382--389,
  2006.

 \bibitem[Shu et al.(2007)]{shu07}
  Shu, F.~H., Galli, D., Lizano, S., Glassgold, A.~E., and Diamond, P.~H.:
  Mean Field Magnetohydrodynamics of Accretion Disks, \apj, 665, 535--553,
  2007.

 \bibitem[Smith et al.(2004)]{smith04}
  Smith, D., Ghosh, S., Dmitruk, P., and Matthaeus, W.~H.: Hall and
  Turbulence Effects on Magnetic Reconnection, \grl, 310, L02805, 2004.

 \bibitem[Speiser(1970)]{speiser70}
  Speiser, T.~W.: Conductivity without collisions or noise, \planss, 18,
  613--1970.

 \bibitem[Stanimirovi{\'c} and Lazarian(2001)]{stanimirovic01}
  Stanimirovi{\'c}, S.~and Lazarian, A.: Velocity and Density Spectra of the
  Small Magellanic Cloud, \apjl, 551, L53--L56, 2001.

 \bibitem[Suresh and Huynh(1997)]{suresh97}
  Suresh, A. and Huynh, H.~T.: Accurate Monotonicity-Preserving Schemes with
  Runge Kutta Time Stepping, \jcp, 136, 83--99, 1997.

 \bibitem[Sweet(1958)]{sweet58}
  Sweet, P.~A.: The Neutral Point Theory of Solar Flares, Electromagnetic
  Phenomena in Cosmical Physics, 6, 123--134, 1958. Conf. Proc. IAU Symposium 6,
  {\em Electromagnetic Phenomena in Cosmical Physics}, ed. B. Lehnert,
  (Cambridge, UK:Cambridge University Press)

 \bibitem[Sych et al.(2009)]{sych09}
  Sych, R., Nakariakov, V.~M., Karlicky, M., and Anfinogentov, S.: Relationship
  between wave processes in sunspots and quasi-periodic pulsations in active
  region flares, \aap, 505, 791--799, 2009.

 \bibitem[Tassis and Mouschovias(2005a)]{tassis05a}
  Tassis, K.~and Mouschovias, T.~C.: Magnetically Controlled Spasmodic
  Accretion during Star Formation. I. Formulation of the Problem and Method
  of Solution, \apj, 618, 769--782, 2005.

 \bibitem[Tassis and Mouschovias(2005b)]{tassis05b}
  Tassis, K.~and Mouschovias, T.~C.: Magnetically Controlled Spasmodic
  Accretion during Star Formation. II. Results, \apj, 618, 783--794, 2005.

 \bibitem[T{\'o}th(2000)]{toth00}
  T{\'o}th, G.: The $\nabla \cdot \mathbf{B}=0$ Constraint in Shock-Capturing
  Magnetohydrodynamics Codes, \jcp, 161, 605--652, 2000.

 \bibitem[Ugai(1992)]{ugai92}
  Ugai, M.: Computer studies on development of the fast reconnection mechanism
  for different resistivity models, \pfb, 4, 2953--2963, 1992.

 \bibitem[Ugai and Tsuda(1977)]{ugai77}
  Ugai, M. and Tsuda, T.: Magnetic field-line reconnexion by localized
  enhancement of resistivity. I - Evolution in a compressible MHD fluid,
  \jpp, 17, 337--356, 1977.

 \bibitem[Vall\'ee(1997)]{vallee97}
  Vall\'ee, J.~P.: Observations of the Magnetic Fields Inside and Outside the
  Milky Way, Starting with Globules (\~{} 1 parsec), Filaments, Clouds,
  Superbubbles, Spiral Arms, Galaxies, Superclusters, and Ending with the
  Cosmological Universe's Background Surface (at \~{} 8 Teraparsecs), \fcp,
  19, 1--89, 1997.

 \bibitem[Vall\'ee(1998)]{vallee98}
  Vall\'ee, J.~P.: Observations of the Magnetic Fields Inside and Outside the
  Solar System: From Meteorites (\~{} 10 attoparsecs), Asteroids, Planets,
  Stars, Pulsars, Masers, to Protostellar Cloudlets (< 1 parsec), \fcp,
  19, 319--422, 1998.

 \bibitem[Wang et al.(2001)]{wang01}
  Wang, X., Bhattacharjee, A., and Ma, Z.~W.: Scaling of Collisionless Forced
  Reconnection, \prl, 87, 265003, 2001.

 \bibitem[Watson et al.(2007)]{watson07}
  Watson, P.~G., Oughton, S., and Craig, I.~J.~D.: The impact of small-scale
  turbulence on laminar magnetic reconnection, \pop, 14, 032301, 2007.

 \bibitem[Yamada(2007)]{yamada07}
  Yamada, M.: Progress in understanding magnetic reconnection in laboratory
  and space astrophysical plasmas, \pop, 14, 058102, 2007.

 \bibitem[Yan et al.(1992)]{yan92}
  Yan, M., Lee, L.~C., and Priest, E.~R.: Fast magnetic reconnection with small shock angles, \jgr, 97, 8277--8293, 1992.

 \bibitem[Zweibel and Yamada(2009)]{zweibel09}
  Zweibel, E.~G.~and Yamada, M.: Magnetic Reconnection in Astrophysical and
  Laboratory Plasmas, \araa, 47, 291--332, 2009.

\end{thebibliography}
\end{document}